\documentclass[structabstract]{aa}  
\usepackage{graphicx}
\usepackage{color}
\usepackage{epsfig}
\usepackage{supertabular}
\usepackage{longtable}
\usepackage{subfigure}
\usepackage{txfonts}
\usepackage{natbib}
\bibpunct{(}{)}{;}{a}{}{,}
\begin{document}
   \title{The early-type dwarf galaxy population of the Hydra\,I cluster\thanks{Based on observations obtained at the European Southern Observatory, Chile (Observing Programmes 065.N--0459(A) and 076.B--0293).}}


   \author{I. Misgeld\inst{1,2} \and S. Mieske\inst{1} \and M. Hilker\inst{1}}


   \institute{European Southern Observatory, Karl-Schwarzschild-Strasse 2, 85748 Garching bei M\"unchen, Germany \\
	\email{imisgeld@eso.org;smieske@eso.org;mhilker@eso.org} \and Argelander Institut f\"ur Astronomie, Universit\"at Bonn, Auf dem H\"ugel 71, 53121 Bonn, Germany\\
              }

   \date{}

 
  \abstract
   {} 
   {We analyse the properties of the early-type dwarf galaxy population ($M_V>-17$ mag) in the Hydra\,I cluster. We investigate the galaxy luminosity function (LF), the colour--magnitude relation (CMR), and the magnitude--surface brightness relation down to $M_V\sim-10$ mag. Another goal of this study is to find candidates for ultra-compact dwarf galaxies (UCDs) in Hydra\,I.}
   {Two spectroscopic surveys performed with Magellan~I/LDSS2 at Las Campanas Observatory and VLT/VIMOS, as well as deep VLT/FORS1 images in $V$ and $I$ bands, covering the central parts of the cluster, were examined. We identify cluster members by radial velocity measurements and select other cluster galaxy candidates by their morphology and low surface brightness. The candidates' total magnitudes and central surface brightnesses were derived from the analysis of their surface brightness profiles. To determine the faint-end slope of the LF, the galaxy number counts are completeness corrected.}
   {We obtain radial velocities for 126 objects and identify 32 cluster members, of which 5 are previously uncatalogued dwarf galaxies. One possible UCD candidate with $M_V=-13.26$ mag is found. Our sample of $\simeq 100$ morphologically selected dwarf galaxies with $M_V>-17$ mag defines a CMR that extends the CMR of the giant cluster galaxies to the magnitude limit of our survey ($M_V\sim-10$ mag). It matches the relations found for the Local Group (LG) and the Fornax cluster dwarf galaxies almost perfectly. The Hydra\,I dwarf galaxies also follow a magnitude--surface brightness relation that is very similar to that of the LG dwarf galaxies. Moreover, we observe a continuous relation for dwarf galaxies and giant early-type galaxies when plotting the central surface brightness $\mu_0$ of a S\'ersic model vs. the galaxy magnitude. The effective radius is found to be largely independent of the luminosity for $M_V>-18$ mag. It is consistent with a constant value of $R_{\mathrm{e}}\sim 0.8$ kpc. We present the photometric parameters of the galaxies as the Hydra\,I Cluster Catalogue (HCC). By fitting a Schechter function to the luminosity distribution, we derive a very flat faint-end slope of the LF ($\alpha = -1.13 \pm 0.04$), whereas fitting a power law for $M_V>-14$ mag gives $\alpha = -1.40 \pm 0.18$.}
   {Our findings of a continuous CMR and $\mu_0$ -- $M_V$ relation for dwarf and giant early-type galaxies suggests that they are the same class of objects. The similarity of those relations to other environments like the LG implies that internal processes could be more important for their global photometric properties than external influences.}


   \maketitle 
%

\section{Introduction}
Dwarf galaxies are the most abundant type of galaxy in the universe. They are most commonly found in galaxy clusters, such as Virgo, Coma and Fornax \citep[e.g.][]{Sandage1984, Binggeli1985, Ferguson1988, Secker1996, Roberts2007}. Also in the Local Group (LG) a large number of dwarf galaxies have been identified \citep[e.g.][and references therein]{Mateo1998, vandenBergh1999, vandenBergh2000, Grebel2003}.

The classification of dwarf galaxies is not standardised in the literature. They are usually distinguished from giant elliptical and spiral galaxies by their low luminosities and low surface brightnesses and, just as for giant galaxies, one refers to early-type and late-type dwarf galaxies (dwarf irregulars). In this paper we focus on \emph{early-type} dwarf galaxies, comprising dwarf elliptical galaxies (dEs) and dwarf spheroidal galaxies (dSphs). In \citet{Grebel2001} dEs are defined as objects with low luminosities ($M_V\gtrsim -17$ mag) and typical central surface brightnesses of $\mu_V \lesssim 21$ mag arcsec$^{-2}$. Dwarf spheroidal galaxies have even lower luminosities ($M_V\gtrsim-14$ mag) and central surface brightnesses ($\mu_V \gtrsim 22$ mag arcsec$^{-2}$). Unless stated otherwise, we use the term dwarf galaxy to refer to both types (dEs and dSphs).

Another probable type of early-type dwarf galaxies has been identified in the nearby galaxy clusters Fornax, Virgo and Centaurus - the so-called ultra-compact dwarf galaxies (UCDs) \citep[e.g.][]{Hilker1999b, Drinkwater2000, Hasegan2005, Jones2006, Mieske2007b}. UCDs are of intermediate nature between dwarf elliptical galaxies and globular clusters in terms of their morphology, being characterised by sizes of $10<r_{\mathrm{eff}}<100$ pc and luminosities between $-13.5<M_V<-11$ mag. The luminosities of the brightest UCDs in Virgo and Fornax are comparable to those of nuclei of dwarf ellipticals and late-type spirals. In order to understand the importance of dwarf galaxies for the evolution of larger galaxies and galaxy clusters, it is essential to investigate their properties in different environments.

Current hierarchical cold dark matter (CDM) models of galaxy formation claim that dwarf-sized galactic fragments were very abundant in the early universe. They represent the building-blocks of larger galaxies. The models predict that Milky Way-sized dark matter haloes as well as cluster-sized haloes contain a large number of dark matter subhaloes \citep{Moore1999}. If every small DM subhalo contained luminous matter, hundreds of dwarf galaxies are expected to be observed in the LG, but already there the number of observed satellite galaxies is too low by about one order of magnitude. This discrepancy is known as the \emph{missing satellites problem} \citep{Klypin1999} and its origin is still a matter of debate. Either the predictions of the hierarchical models are not reliable, or there is a large number of luminous satellites being not yet discovered, or a major fraction of low-mass DM haloes has not formed any stars. In fact, the analysis of SDSS data recently led to the discovery of several extremely faint ($M_V\gtrsim -8$ mag) LG dwarf galaxies \citep[e.g.][]{Willman2005a, Willman2005b, Zucker2006a, Zucker2006b, Belokurov2006, Irwin2007, Simon2007}, proving that such hardly detectable objects do exist. But they are still not abundant enough to explain the expected DM sub-structure.

\subsection{The galaxy luminosity function}
A well known way to quantify the discrepancy between the number of observed and the number of predicted dwarf galaxies is the determination of the galaxy luminosity function (LF). An analytic function for the LF was proposed by \citet{Schechter1976} as
\[ \phi(L)dL =\phi^* \left(\frac{L}{L^*}\right)^{\alpha} e^{-(L/L^*)} d\left(\frac{L}{L^*}\right).\]
The logarithmic faint-end slope $\alpha$ can be contrasted with the predicted slope of $\simeq-2$ for the mass spectrum of cosmological dark-matter haloes \citep[e.g.][]{Press1974, Moore1999, Jenkins2001}. Inconsistent with the predictions, rather flat slopes of $-1.0 \lesssim \alpha \lesssim -1.5$ have been derived from studies of various environments ranging from the LG to the Coma cluster \citep[e.g.][]{vandenBergh1992, Trentham2002, CZ2003, Hilker2003, Mobasher2003, Trentham2005, Chiboucas2006, Mieske2007, Rines2007, Penny2008}. Other authors, however, reported on steeper slopes \citep[e.g.][and references therein]{Sabatini2003, Milne2007}. The LF of the Hydra\,I cluster has previously been determined by \citet{Yamanoi2007} who find $\alpha \sim -1.6$ in the magnitude range $-20<M_{B,R}<-10$. This is a steeper slope than reported by \citet{Yagi2002} who give $\alpha=-1.31$ for the faint end slope of the composite LF of 10 nearby clusters (including Hydra\,I) at $-23<M_R<-16$. \citet{CZ2003} find $\alpha=-1.21$ for the composite LFs of six clusters (also including Hydra\,I) at $-22<M_R<-14$, derived from deep spectroscopic samples.

\subsection{Photometric scaling relations for early-type galaxies}
Correlations among global parameters of early-type galaxies can provide insight into the physical processes that have impact on the formation mechanisms and the evolution of these galaxies. For example, luminosity, colour, surface brightness, central velocity dispersion and the $\mathrm{Mg}_2$ absorption line index are related to each other \citep[e.g.][]{Faber1976, Kormendy1977, Kormendy1985, Djorgovski1987, Ferguson1988, Bender1992, Bernardi2003b, Bernardi2003c, Bernardi2003d, Chang2006}. In particular, the colour--magnitude relation and the magnitude--surface brightness relation connect the physical properties of the underlying stellar population and the structural properties with the galaxy masses. Investigating those scaling relations in multiple environments sets constraints for galaxy formation and evolutionary models of early-type galaxies.

\subsubsection{Colour--magnitude relation}
\label{sec:cmrintro}
A tight colour--magnitude relation (CMR) for early-type cluster galaxies has been known for a long time \citep{Visvanathan1977}. It is most commonly explained by an increase of the mean stellar metallicity with increasing galaxy mass, leading to redder colours of the more luminous galaxies \citep[e.g.][]{Kodama1997, Ferreras1999}. With a large sample from the Sloan Digital Sky Survey, \citet{Gallazzi2006} confirmed this perception. The metallicity of a galaxy strongly depends on the fraction of gas that has been turned into stars. Due to their deeper potential well, massive galaxies are able to retain their interstellar gas and stellar ejecta longer and more effectively than low-mass galaxies since the escape velocity of stellar yields depends on the galaxy mass. Hence, subsequent generations of stars will be formed out of already enriched gas. Because giant elliptical galaxies with their high star formation rate (SFR) consume their gas very fast, the mean stellar metallicity can reach high values in a short time. That accounts for the reddest colours of the most luminous galaxies.

\citet{Koeppen2007} demonstrated that a variable integrated galactic initial mass function (IGIMF) that depends on the SFR can also explain the observed mass--metallicity relation. For a low SFR the IGIMF differs from the standard IMF in the sense that fewer high mass stars are expected to form. It has been found that galaxies with a low current SFR contain star clusters with a lower maximum mass \citep{Weidner2004}, and less massive star clusters less likely contain very massive stars \citep{Kroupa2003}. Since less massive galaxies are expected to have lower star formation rates, the effective upper mass limit for stars in such galaxies is lower. This reduces the number of type II supernovae as the main source of $\alpha$-elements. Hence, the dependence of the IGIMF on the SFR implies a dependence of the metal abundance on the mass of the galaxy.

As an alternative to the aforementioned scenarios, \citet{Worthey1994} and \citet{Poggianti2001} suggested that a change in the mean age of a stellar population could at least in part explain the CMR. A stellar population will gradually redden as stars with increasing age evolve off the main sequence towards the red giant branch. In this picture, redder colours of more massive galaxies imply systematically older ages. Observational evidence for this was given by \citet{Rakos2004}, who reported on younger ages of dwarf elliptical galaxies in the Coma and Fornax clusters. Hence, an increase of age at fixed metallicity has the same effect on galaxy colours as an increase of metallicity at fixed age. This ambiguous interpretation of the CMR is based on the well-known \emph{age-metallicity degeneracy} of integrated optical colours.

\subsubsection{Magnitude--surface brightness relation}
It has been known for decades that elliptical galaxies follow a relation between effective radius and absolute magnitude \citep{Kormendy1977}. Dwarf elliptical galaxies, on the other hand, follow a $r_{\mathrm{eff}}$-$M_V$ relation unequal in slope to that of brighter giant ellipticals \citep{Bender1992}. It was believed that a dichotomy between dwarf elliptical and giant elliptical galaxies also exists in the magnitude--surface brightness plane (based on exponential and de Vaucouleurs fits). According to this, dwarf galaxies follow a relation in the sense that brighter dEs have higher surface brightnesses, whereas the surface brightness of giant elliptical galaxies decreases with increasing luminosity \citep[e.g.][]{Kormendy1985, Ferguson1988, Bender1992}.

\citet{Graham2003}, however, showed that there does not exist a dichotomy when relating the magnitude $M$, the profile shape index $n$ and the central surface brightness $\mu_0$ of \citet{Sersic1968} profile fits with each other (except maybe for the most luminous giant ellipticals that deviate from a smooth relation due to their low surface brightness cores). In this study, the interpretation of different galaxy formation mechanisms between giant and dwarf galaxies is not supported.

\citet{Phillipps1988} and \citet{Irwin1990} argued against the existence of a magnitude--surface brightness relation for dEs. They claimed that the relation in eye-selected samples of dwarf galaxies is merely a result of selection effects, but \citet{Karick2003} and \citet{Mieske2007} observed such a relation for Fornax dEs for which cluster membership was assured either by an un-biased radial velocity survey (Karick) or by SBF measurements (Mieske).

\subsection{The Hydra\,I galaxy cluster (Abell 1060)}
\begin{figure}
	\resizebox{\hsize}{!}{\includegraphics{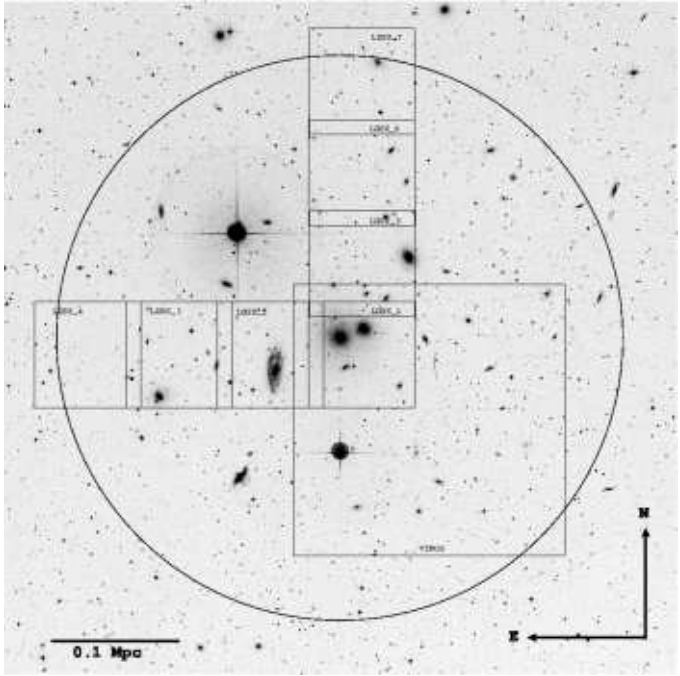}}
	\caption{$45'\times 45'$ ($540 \times 540$ kpc at the cluster distance) image of the Hydra\,I cluster centred on NGC 3311, extracted from the Digital Sky Survey. The small squares are the fields observed with LDSS2. The large square marks the VIMOS-pointing. The circle indicates the cluster core radius $r_{\mathrm{c}}=170\ h^{-1}$ kpc \citep{Girardi1995}, adopting $h=0.75$.}
	\label{fig:hydracluster}
\end{figure}

Hydra\,I is a relatively poor cluster [richness class $R=1$ \citep{Abell1958}, BM classification III \citep{Bautz1970}]. A pair of bright galaxies is located near the cluster centre. NGC 3309 is a regular giant elliptical galaxy (E3). The brighter and larger galaxy NGC 3311 possesses an extended cD halo and an extraordinarily rich globular cluster system \citep{vandenBergh1977, McLaughlin1995, Mieske2005, Wehner2008}.

The cluster is the prototype of an evolved and dynamically relaxed cluster, being dominated by early-type galaxies and having a regular core shape. From X-ray measurements, \citet{Tamura2000} derived  an isothermal distribution of the intracluster medium within $\sim 160\ h^{-1}\ \mathrm{kpc}$. They give $2.1 \times 10^{14}\ h^{-1}\ \mathrm{M_{\sun}}$ as the cluster virial mass. From optical studies, \citet{Girardi1995} found a core radius of $r_{\mathrm{c}} = 170\ h^{-1}\ \mathrm{kpc}$. Applying the virial theorem to the member galaxies, \citet{Girardi1998} calculated a virial mass of $1.9 \times 10^{14}\ h^{-1}\ \mathrm{M_{\sun}}$.

\citet{Mieske2005} estimated the distance to Hydra\,I from $I$-band surface brightness fluctuations (SBF). They found the distance to be $41.2 \pm 1.4\ \mathrm{Mpc}$ (distance modulus $(m-M)=33.07 \pm 0.07\ \mathrm{mag}$). This is at the low end of distance estimates by other authors whose mean is $\sim 15$\% higher \citep[see discussion in][]{Mieske2005}. From a deep spectroscopic sample of cluster galaxies, extending to $M_R=-14\ \mathrm{mag}$, \citet{CZ2003}, hereafter CZ03, derived $\overline{cz}=3683\pm46\ \mathrm{km\ s^{-1}}$ as the mean cluster redshift with a velocity dispersion of $\sigma = 724\pm 31\ \mathrm{km\ s^{-1}}$. This corresponds to a distance of $51.2 \pm 5.7\ \mathrm{Mpc}$, assuming $H_0=72\pm 8\ \mathrm{km\ s^{-1}\ Mpc^{-1}}$ \citep{Freedman2001}. Already earlier, it was discovered that the cluster is clearly isolated in redshift space, having no foreground galaxies and no background galaxies up to $cz\sim 8000\ \mathrm{km\ s^{-1}}$ \citep{Richter1982, Richter1987}. This implies huge empty regions of space of about 50 Mpc path length in front and behind the cluster.

The Hydra\,I cluster is close enough for current 8-m class telescopes like the VLT to resolve faint dwarf galaxies under good seeing conditions. Only a few images of about $7\times7$ arcmin (the typical field-of-view size for most CCD cameras) are needed to observe the cluster centre as well as areas out to about one core radius (see Fig. \ref{fig:hydracluster}).

In this paper we present a study of the early-type dwarf galaxy population in the Hydra\,I cluster. The study is based on two spectroscopic surveys and a photometric survey. One spectroscopic survey was performed with LDSS2 at Magellan~I and aimed at searching for normal cluster dwarf galaxies. The other spectroscopic survey -- performed with VLT/VIMOS -- focused on the search for UCDs. The photometric survey, based on deep VLT/FORS1 images, was used to identify other cluster dE/dSph candidates. The paper is organised as follows. In Sect. \ref{sec:spectroscopy} we describe the observations, the candidate selection and the data reduction for the two spectroscopic surveys. In Sect. \ref{sec:photometry} the photometric analysis of the candidate cluster dEs/dSphs, as selected from the VLT/FORS1 images, is addressed. Our results are presented in Sect. \ref{sec:results}. We summarise and discuss our findings in Sect. \ref{sec:discussion}.

\section{Spectroscopy}
\label{sec:spectroscopy}

\subsection{Observations and selection of candidates}

\subsubsection{The dwarf galaxy sample}
For the first spectroscopic survey, seven fields in the central region of the Hydra\,I cluster were observed with Magellan I at Las Campanas Observatory together with the Low Dispersion Survey Spectrograph (LDSS2) in April/May 2001 (see Fig. \ref{fig:hydracluster}). The goal of the LDSS2-survey was to identify cluster dwarf galaxies by radial velocity measurements.

On Magellan~I, LDSS2 images a $7.5'$ diameter field onto the LCO SITe\#1 detector of $2048\times 2048$ pixel with a scale of $0.378''$/pixel. The high dispersion grism with a central wavelength of 4200\AA~and a dispersion of 2.4 \AA/pixel was used. With a slit width of $1.25''$ ($\approx 3$ pixel), the effective resolution was about 7 \AA, corresponding to 525 km s$^{-1}$ at 4000 \AA. Except for fields 4, 6 and 7 (see Fig. \ref{fig:hydracluster}), two slit masks were observed in each field. For each mask two exposures were taken, each with an integration time of 1200 s.

\begin{figure}
	\resizebox{\hsize}{!}{\includegraphics{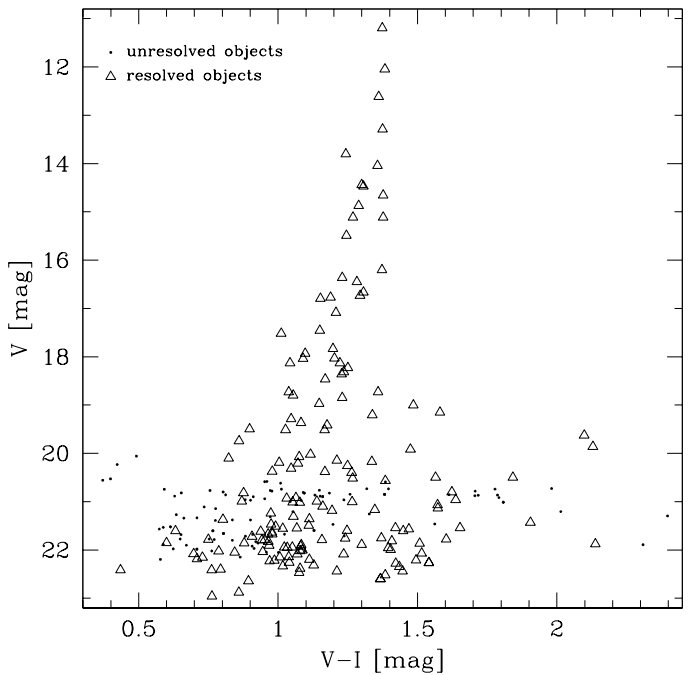}}
	\caption{Colour--magnitude diagram of all observed objects in the seven LDSS2 fields (the dwarf galaxy sample). Dots (open triangles) mark unresolved (resolved) sources, according to SExtractor star-galaxy classifier \citep{Bertin1996}. The photometry is taken from VLT/FORS1 images (see Sect. \ref{sec:photobservation}).}
	\label{fig:cmdselectldss}
\end{figure}

The objects observed with LDSS2 were selected from VLT/FORS1 images (see Sect. \ref{sec:photobservation}). This dwarf galaxy sample contains both the bright giant elliptical galaxies and a number of possible dwarf elliptical galaxies (dE, dE,N, dS0) or dwarf spheroidal galaxies (dSph), selected by their morphology. Compact, unresolved objects, being candidates for globular clusters (GCs), isolated nuclei from dissolved dEs, or ultra-compact dwarf galaxies (UCDs), complement the sample. Fig. \ref{fig:cmdselectldss} shows a colour--magnitude diagram of all observed objects.

\subsubsection{The UCD sample}
The second spectroscopic survey was performed with the aim of searching for UCDs in the Hydra\,I cluster. The observations were carried out with VLT/VIMOS at ESO/Paranal in February 2006 [ESO observing programme 076.B-0293]. Four quadrants of $7\arcmin \times 8\arcmin$ were observed. The pixel scale was $0.205''$/pixel. Multi-slit masks and the medium resolution MR grism with a wavelength coverage of [4800:10\,000]~\AA~and a dispersion of 2.5 \AA/pixel were used. With a slit width of $0.8''$ ($\approx 4$ pixel) the effective resolution was about 10~\AA, corresponding to 750~km~s$^{-1}$ at 4000~\AA. The total integration time was 4200 s, subdivided into two exposures. Fig. \ref{fig:hydracluster} shows the observed regions overlaid to a DSS image of the cluster.

\begin{figure}
	\resizebox{\hsize}{!}{\includegraphics{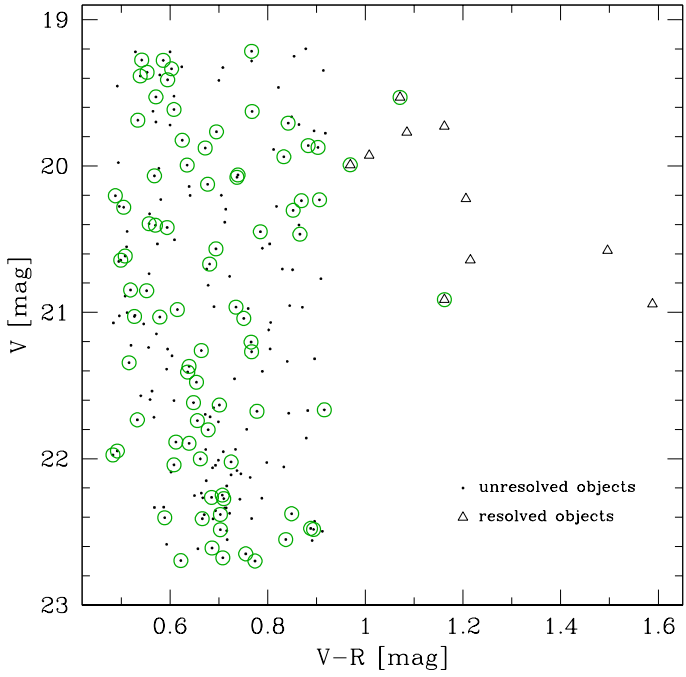}}
	\caption{Colour--magnitude diagram of all photometrically selected objects in the VIMOS-pointing (the UCD sample). Dots are unresolved sources and open triangles are dwarf galaxy candidates. Green open circles mark objects for which a slit could be allocated on the masks.}
	\label{fig:cmdselectvimos}
\end{figure}

The candidates for spectroscopy with VIMOS were selected from $V$ and $R$ pre-images. The UCD sample is restricted in magnitude and colour to $19.2<V<22.7$ and $0.48<V-R<0.93$. Most of the selected objects are GC/UCD candidates, hence, unresolved sources. Additionally, 10 dwarf galaxy candidates were included in the sample. A colour--magnitude diagram of the photometrically selected objects and the subset of finally observed objects in the VIMOS-pointing is shown in Fig. \ref{fig:cmdselectvimos}.

\subsection{Data reduction and radial velocity measurements}

For the LDSS2 data set, the standard process of data reduction, comprising bias subtraction, cosmic ray removal by means of the \texttt{lacosmic} routine \citep{vanDokkum2001}, correction for spatial distortion, flatfield normalisation and wavelength calibration, was performed with the IRAF-packages \texttt{onedspec} and \texttt{twodspec}. After these reduction steps, the one-dimensional object spectra were extracted with simultaneous sky subtraction. The VIMOS-spectra were extracted by using the data reduction pipeline for VIMOS as provided by ESO's Data Flow System Group. This pipeline performs the basic data reduction steps mentioned above in an automated manner.

Radial velocities were determined by performing Fourier cross-correlations between object and template spectra, using the IRAF-task \texttt{fxcor} in the \texttt{rv} package. The object spectra were initially cross-correlated against six template spectra. One template is a galaxy spectrum of NGC 1396 \citep{Dirsch2004}, a second one is HD 1461, an old metal-rich star in the solar neighbourhood \citep{Chen2003}. The remaining four template spectra were taken from \citet{quintana1996}. Three of them are spectra from early-type galaxies (NGC 1407, NGC 1426, NGC 1700) and one is a synthetic template. The wavelength range of the three galaxy spectra [3800:6500]~\AA~is similar to that of the object spectra. Only the wavelength range of the synthetic spectrum extends to about 7400~\AA, making it suitable for cross-correlating the VIMOS-spectra.

The four template spectra from \citet{quintana1996} were found to give the best cross-correlation results. The peak of the CCF was most pronounced and the coefficient $R$, which gives the significance of the cross-correlation match \citep{tonry1979}, was similar for all four templates. Also the four obtained radial velocities were consistent within the errors. The $R$-values of the correlations with the other two templates were clearly lower (by about 20--50\%).

\begin{figure}
	\subfigure{\includegraphics[width=4.4cm]{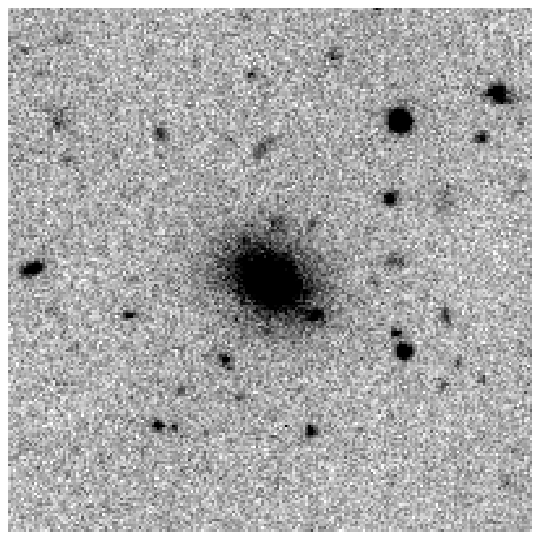}}
	\subfigure{\includegraphics[width=4.4cm]{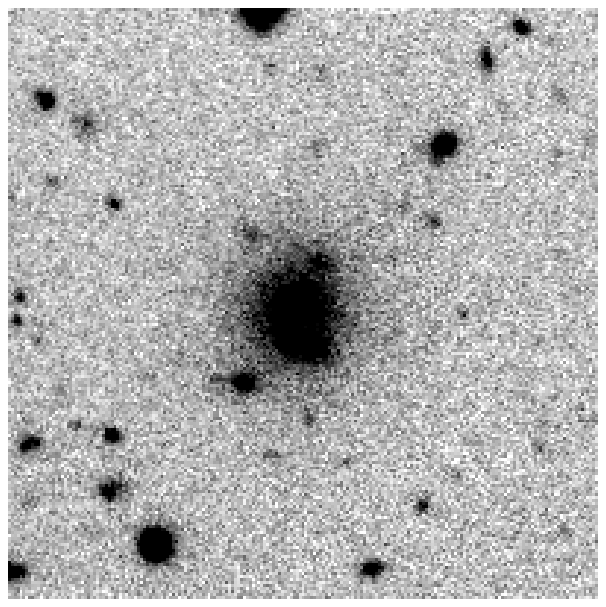}}\\
	\subfigure{\includegraphics[width=4.4cm]{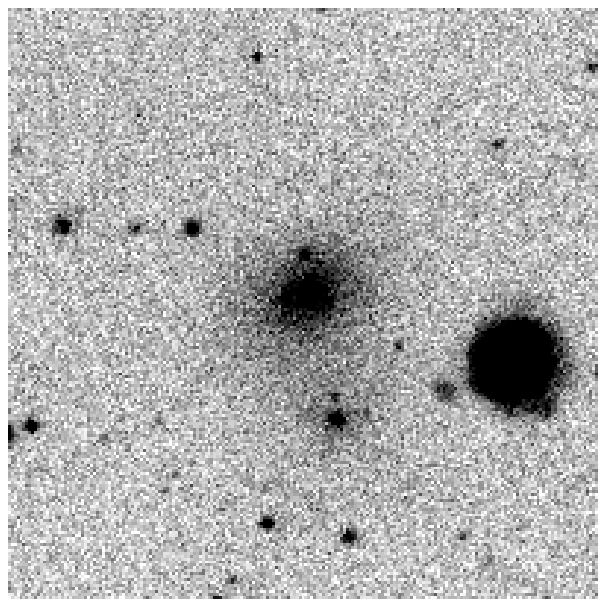}}
	\subfigure{\includegraphics[width=4.4cm]{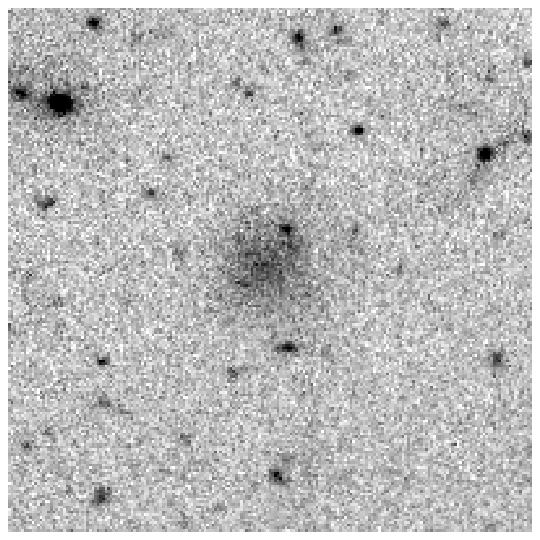}}
	\caption{Thumbnails of four cluster dwarf galaxy candidates that fulfil the selection criteria. The thumbnail sizes are $40''\times40''$ ($8\times8$ kpc at the cluster distance).}
	\label{fig:thumbnails}
\end{figure}

We used the $R$ coefficient as an indicator for the reliability of our results. Only values of $R\geq4$ (averaged over the four templates) were considered reliable. This was the limit where the peak of the CCF could easily be distinguished from the noise. Several correlations still showed a clear peak in the CCF, despite having a relatively low S/N ratio ($R<4$). If in those cases the CCF-peak was visible within the errors at the same radial velocity for all four cross-correlations, the measurement was accepted.

\section{Photometry of early-type dwarf galaxies}
\label{sec:photometry}

\subsection{Observations and selection of dE/dSph candidates}
\label{sec:photobservation}
The imaging data for Hydra\,I were obtained in a VLT/FORS1 service mode run in April 2000 at ESO/Paranal [ESO observing programme 65.N-0459(A)]. Seven $7'\times 7'$ fields were observed in Johnson $V$ and $I$ filters. All images were taken during dark time with a seeing between $0.5''$ and $0.7''$. The integration time was $3\times 8$ min for the $V$ images and $9\times 5.5$ min for the $I$ images. Since most of the bright cluster galaxies were saturated on the long exposures, their photometric parameters were determined from unsaturated short integration time images (30 sec in $V$ and~$I$).

In analogy to our investigations in the Fornax cluster \citep{Hilker2003, Mieske2007}, our strategy for identifying dwarf galaxy candidates is a combination of visual inspection of the images and the use of SExtractor \citep{Bertin1996} detection routines. In a first step, we simulated several LG dEs and dSphs (projected to the cluster distance) and added them to the images. The photometric parameters for these simulated galaxies were taken from \citet{Grebel2003}. After that, the images were inspected by eye and we selected candidate cluster dwarf galaxies by means of their morphological resemblance to the simulated LG dwarfs. The main criterion was that they showed an extended low surface brightness envelope and no substructure or clear features such as spiral arms. The search resulted in the selection of 73 previously uncatalogued dE/dSph candidates. Fig. \ref{fig:thumbnails} shows $V$-band images of four cluster dwarf galaxy candidates.

We then used the SExtractor detection routines with the aim of quantifying the detection completeness in our data and in order to find more dwarf galaxy candidates, especially at the faint magnitude and surface brightness limits. The detection-sensitive parameters of SExtractor were optimised such that most of the objects of the by-eye-catalogue were detected by the programme. Only three of the obvious by-eye detections were not found by SExtractor, due to the proximity of a bright foreground star. We searched for new dwarf galaxy candidates by focusing on those sources in the SEXtractor output catalogue whose photometric parameters matched the parameter range of the simulated dE galaxies. In Sect. \ref{sec:lfunction} we discuss in detail the cuts in \texttt{magbest}, \texttt{mupeak}, \texttt{fwhm} and \texttt{area} that were applied to constrain the output parameter space to those found for the simulated dEs. A total of 9 additional SExtractor detections, covering a magnitude range $-11.7 < M_V < -9.7$ mag, were found and added to the by-eye catalogue.

\begin{figure}
	\resizebox{\hsize}{!}{\includegraphics{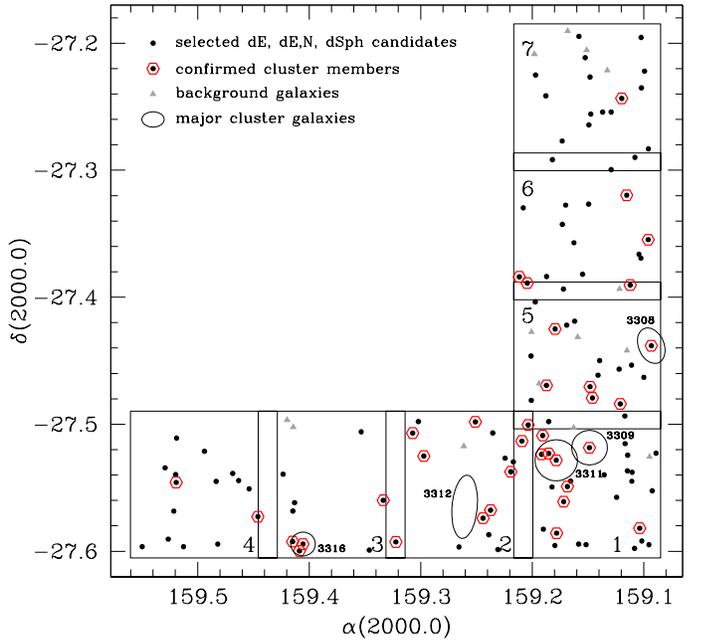}}
	\caption{Coordinate map of the cluster dwarf galaxy candidates, confirmed cluster members, background galaxies and major cluster galaxies with their corresponding NGC number. The seven observed VLT/FORS1 fields are marked by large open squares, the field number is indicated.}
	\label{fig:fields}
\end{figure}

Finally, 36 spectroscopically confirmed cluster early-type galaxies from our study and from CZ03 were added to the photometric sample. The CZ03 catalogue has a full spatial coverage over the observed fields and a limiting magnitude of $M_R=-14$ mag. For comparison, also 14 identified background galaxies from the LDSS2 survey were added to the sample. A map of the observed fields and the 132 objects chosen for the photometric analysis is presented in Fig. \ref{fig:fields}.

\subsection{Data analysis}
The surface brightness profile for each selected object was derived by fitting elliptical isophotes to the galaxy images, using the IRAF-task \texttt{ellipse} in the \texttt{stsdas}\footnote{Space Telescope Science Data Analysis System, STSDAS is a product of the Space Telescope Science Institute, which is operated by AURA for NASA.} package. Sky subtraction and isophote fitting for each object were performed on cut out thumbnails that extended well into the sky region (see Fig. \ref{fig:thumbnails}). Isophotes with fixed centre coordinates, ellipticity and position angle were fitted to the galaxy images. In particular for the brightest cluster galaxies ($V\lesssim 16$ mag), the ellipticity considerably changed from the inner to the outer isophotes. In those cases the ellipticity was not fixed during the fitting procedure.

The photometric parameters of the objects were derived from the analysis of their surface brightness profiles: the total magnitude from a curve of growth analysis and the central surface brightness from both an exponential and a S\'ersic fit to the profile. For the fit we excluded the inner $1''$ (about 1.5 seeing disks) and the outermost part of the profile, where the measured surface brightness is below the estimated error of the sky background. Photometric zero points were taken from \citet{Mieske2005}. In order to correct for interstellar absorption and reddening the values from \citet{Schlegel1998} were used. They give $A_V=0.263$ mag and $E(V-I)=0.110$ mag for the coordinates of the Hydra\,I cluster. Table \ref{tab:calcoeff} in the appendix lists the photometric calibration coefficients for the observed fields (only available on-line).

The obtained photometric parameters along with the available radial velocities for all early-type galaxies in our sample (111 objects) are presented as the Hydra\,I Cluster Catalogue (HCC) -- see Table \ref{tab:hydrasample} in the appendix (only available on-line).

\section{Results}
\label{sec:results}
\begin{figure}
	\resizebox{\hsize}{!}{\includegraphics{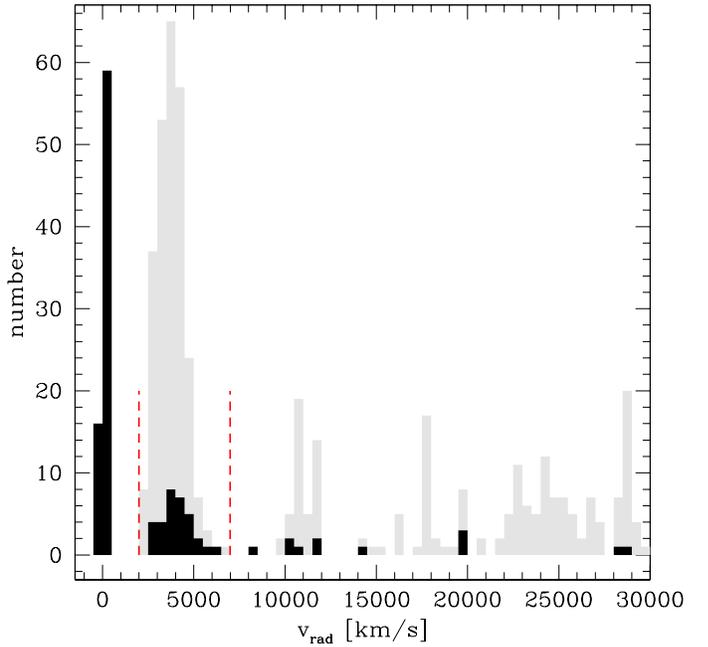}}
	\caption{Radial velocity distribution of all successfully measured objects with $v_{\mathrm{rad}}<30\,000$ km s$^{-1}$ (black histogram). The grey histogram shows the data from CZ03. Vertical dashed lines mark the velocity range assumed for cluster membership. Note also the overdensity at $v_{\mathrm{rad}} \sim 11\,000$ km s$^{-1}$ from which we identify five members.}
	\label{fig:vrad}
\end{figure}

In this section a detailed analysis of the spectroscopic and photometric data is presented. The results of the spectroscopic surveys are given in Sect. \ref{sec:specresult}. The properties of the newly discovered GC/UCD candidates are summarised in Sect. \ref{sec:UCDGC}. In Sects. \ref{sec:hydracmd} and \ref{sec:magmurel} the colour--magnitude and the magnitude--surface brightness relation of the Hydra\,I dwarf galaxies are presented. Section \ref{sec:lfunction} addresses the galaxy luminosity function for early-type dwarf galaxies in Hydra\,I.

\subsection{Spectroscopic samples}
\label{sec:specresult}
\begin{figure}
	\resizebox{\hsize}{!}{\includegraphics{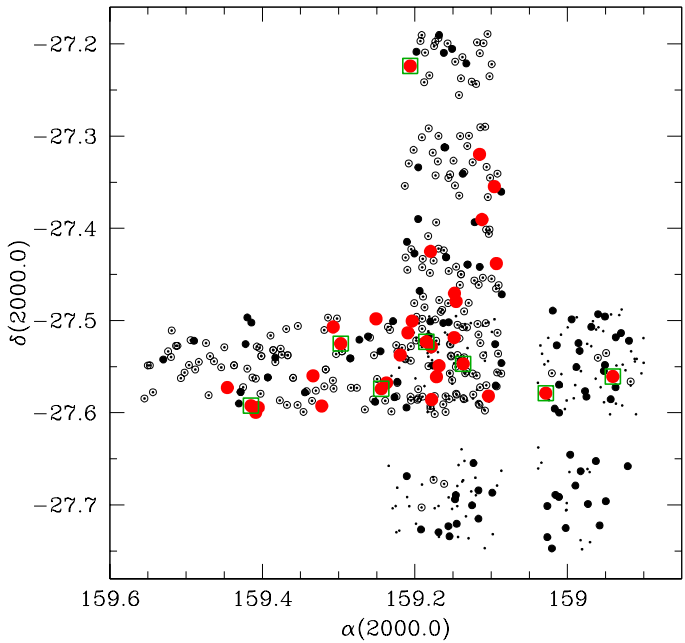}}
	\caption{Coordinates of the 494 objects that were selected for spectroscopy (small dots). Small open circles mark the 365 observed objects, small filled circles are the 126 objects for which a reliable radial velocity could be determined. Red filled circles are cluster members, green open squares mark the 8 newly confirmed cluster members.}
	\label{fig:speccoo}
\end{figure}

A total of 494 objects were selected for spectroscopy. For 365 of these a slit could be allocated on the masks. Reliable radial velocities were measured for 126 objects. Fig. \ref{fig:speccoo} shows a coordinate map of all objects that were selected for spectroscopy. The criterion for cluster membership was adopted to be $2000<v_{\mathrm{rad}}<7000$ km~s$^{-1}$ (see Fig. \ref{fig:vrad}). Radial velocity uncertainties were of the order of 20--100 km~s$^{-1}$. Altogether, 75 foreground stars, 19 background objects and 32 cluster members were identified from both spectroscopic samples. 24 of the confirmed cluster galaxies are already known from CZ03, five are previously uncatalogued dwarf galaxies. Three compact cluster members are candidates for bright GCs and/or UCDs.

\subsubsection{The dwarf galaxy sample}
\begin{figure}
	\resizebox{\hsize}{!}{\includegraphics{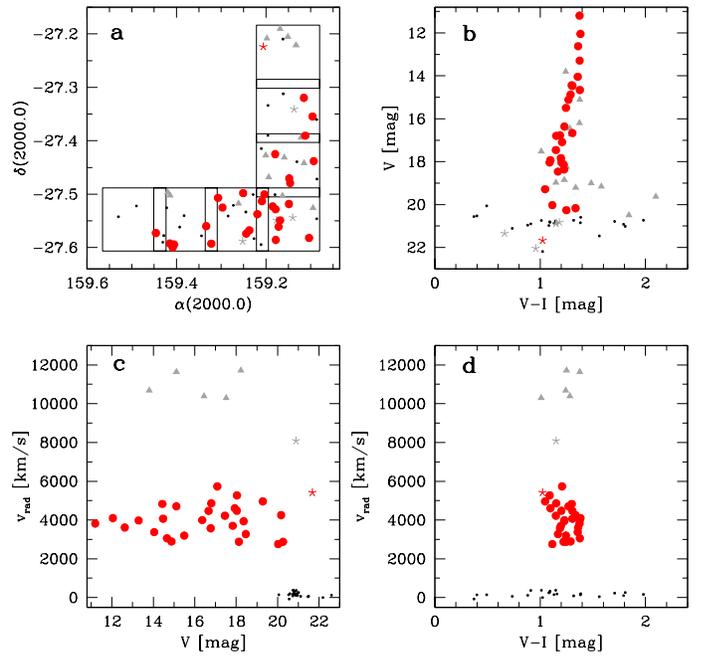}}
	\caption{Properties of all successfully observed objects in our dwarf galaxy sample. Filled red circles are confirmed cluster members, grey triangles represent resolved background objects. Small black dots are foreground stars, asterisks are unresolved cluster members (red) or background objects (grey). \textbf{a)} Coordinate map of all foreground stars, background sources and cluster members. The observed fields are indicated by large open squares. \textbf{b)} Colour--magnitude diagram of all objects. \textbf{c)} Magnitude velocity diagram of all objects, except for background objects with $v_{\mathrm{rad}}>13\,000$ km s$^{-1}$. \textbf{d)} Colour velocity diagram of all objects, except for background objects with $v_{\mathrm{rad}}>13\,000$~km~s$^{-1}$.}
	\label{fig:ldssprops}
\end{figure}

279 objects were observed with LDSS2. Reliable radial velocities of 71 objects could be derived. 24 foreground stars, 18 background objects and 29 objects belonging to the cluster were identified. Fig. \ref{fig:ldssprops} gives a coordinate map, a colour--magnitude diagram, a magnitude velocity and a colour velocity diagram of the successfully observed objects.

In Table \ref{tab:hydrasample} the radial velocities of the 28 resolved cluster members ($V<20.5$ mag) are given. The mean value is $\bar{v}_{\mathrm{rad}} = 3982 \pm 148$ km s$^{-1}$ with a standard deviation of $\sigma=784$ km s$^{-1}$. This deviates by $2\sigma$ from the result of CZ03. However, there is no systematic velocity shift in the LDSS2 data. For the 24 previously known galaxies, differences of $-110<\Delta v_{\mathrm{rad}}<140$ km s$^{-1}$ to the radial velocities from CZ03 are measured. Taking into account the relatively large velocity error of $\pm 80$ km s$^{-1}$ in the literature values, 75\% of all LDSS2 velocities are still consistent with them. Furthermore, the mean radial velocity of the 24 galaxies from the CZ03 catalogue, which are also in our sample, is $3998 \pm 150$ km s$^{-1}$, in agreement with our result. The large discrepancy between the mean radial velocity of our dwarf galaxy sample and the whole CZ03 sample can thus be explained by selection effects.

We note that there is no significant difference in velocity dispersion between brighter (thus more massive) cluster galaxies ($V<16$ mag) and fainter ones, which would be an indication of mass segregation. We find $\sigma=634^{+376}_{-165}$ km s$^{-1}$ for the brighter galaxies and $\sigma=862^{+268}_{-190}$ km s$^{-1}$ for the fainter ones (90\% confidence level). The number counts are too low to judge on a possible larger velocity dispersion for dwarf galaxies as found in the Fornax cluster \citep{Drinkwater2001}.

The colour--magnitude diagram in Fig. \ref{fig:ldssprops}b clearly shows the CMR of cluster galaxies with $V<21$ mag, in the sense that fainter galaxies are on average bluer. The only cluster member with $V>21$ mag is an unresolved source, hence a GC/UCD candidate (see Sect. \ref{sec:UCDGC}). The cluster galaxies have colours of $1.02<V-I<1.38$. Five objects ($V<18.3$ mag) scattering around the CMR belong to a background group at $v_{\mathrm{rad}} \sim 11\,000$ km s$^{-1}$ (see also Fig. \ref{fig:vrad}). The cluster CMR is analysed in more detail in Sect. \ref{sec:hydracmd}.

\begin{figure}
	\resizebox{\hsize}{!}{\includegraphics{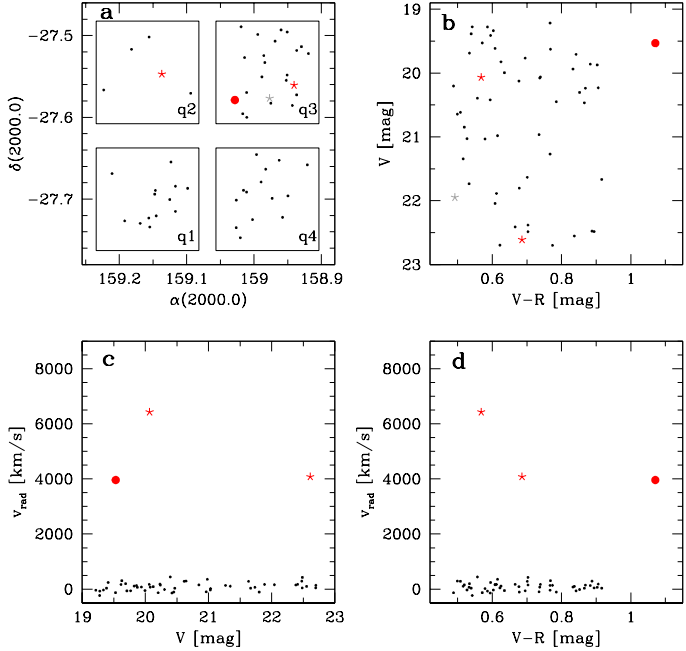}}
	\caption{Properties of all successfully observed objects in the UCD sample. The symbols are as in Fig. \ref{fig:ldssprops}. \textbf{a)} Coordinate map. The four quadrants (q1--q4) of the VIMOS pointing are indicated by large open squares. Note the low number of successfully observed objects in q2 due to the mask misalignment. \textbf{b)} Colour--magnitude diagram of all objects. \textbf{c)} Magnitude velocity diagram of all objects, except for background objects with $v_{\mathrm{rad}}>9000$ km s$^{-1}$. \textbf{d)} Colour velocity diagram of all objects, except for background objects with $v_{\mathrm{rad}}>9000$ km s$^{-1}$.}
	\label{fig:vimosprops}
\end{figure}

\subsubsection{The UCD sample}
Radial velocities of 55 objects could be determined in the UCD sample. 51 foreground stars, one background object and three cluster members were identified. One of the cluster members is a dE galaxy, the others are GC/UCD candidates. Fig. \ref{fig:vimosprops} shows the properties of the successfully observed objects in the same way as Fig. \ref{fig:ldssprops} does for the dwarf galaxy sample. It is obvious from Fig. \ref{fig:vimosprops}a that there is a very low number of successfully observed objects in the second VIMOS-quadrant (q2). Only 5 out of 27 possible spectra could be extracted due to a misalignment of the slit mask in this quadrant. Table \ref{tab:vimosobserv} compares the number of selected and observed objects with the number of objects for which a radial velocity could be determined. Overall, 55 out of 215 photometrically selected compact object candidates were successfully observed, corresponding to a completeness of 26\%.

\begin{table}
	\caption{Objects in the VIMOS-pointing. (A) number of selected objects, (B) number of observed objects, (C) number of objects for which a radial velocity could be determined (see also Fig. \ref{fig:vimosprops}a).}
	\label{tab:vimosobserv}
	\centering
	\begin{tabular}{cccc}
	\hline\hline
	Quadrant & (A) & (B) & (C) \\
	\hline
	q1 & 43 & 16 & 13 \\
	q2 & 85 & 27 & 5 \\ 
	q3 & 62 & 27 & 23 \\ 
	q4 & 25 & 16 & 14 \\
	\hline
	$\Sigma$ & 215 & 86 & 55 \\
	\hline
	\end{tabular}
\end{table}

\subsection{GC/UCD candidates in Hydra\,I}
\label{sec:UCDGC}

\begin{table*}
	\caption{Properties of the GC/UCD candidates found in both samples. The absolute magnitude $M_V$ is calculated by adopting a distance modulus of $(m-M)=33.07$ mag \citep{Mieske2005}.}
	\label{tab:compactobj}
	\centering
	\begin{tabular}{lrrccccc}
	\hline\hline
	ID & $\alpha$(2000.0) & $\delta$(2000.0) & $v_{\mathrm{rad}}$ [km s$^{-1}$] & $V$ [mag] & $V-I$ [mag] & $V-R$ [mag] & $M_V$ [mag] \\
	\hline
	7a-31&	10:36:49.47&	-27:13:26.26&	$5417\pm 52$ & 21.67 & 1.02 & &  -11.66\\
	q2-15&	10:36:32.82&	-27:32:49.20&	$6429\pm 55$ & 20.07 & & 0.57 & -13.26\\
	q3-18&  10:35:45.71&	-27:33:38.41&	$4079\pm 66$ & 22.61 & & 0.69 & -10.72 \\
	\hline
	\end{tabular}
\end{table*}

We identify three compact unresolved sources as cluster members. Table \ref{tab:compactobj} gives the radial velocity and the photometric properties of those objects. They all have absolute magnitudes brighter than $\omega$ Centauri (NGC 5139), the brightest Milky Way globular cluster (\citet{Harris1996} gives $M_V=-10.29$ mag for $\omega$ Cen). Hence, they either represent very bright globular clusters, or they belong to the class of the so-called ultra-compact dwarf galaxies (UCDs) that were discovered during the past few years in the nearby galaxy clusters Fornax, Virgo and Centaurus \citep[e.g.][]{Hilker1999b, Drinkwater2000, Hasegan2005, Jones2006, Mieske2007b}.

Given its absolute magnitude of $M_V=-13.26$ mag, the brightest object in this sample clearly falls into the UCD magnitude range of $-13.5<M_V<-11.0$ mag. With $v_{\mathrm{rad}}=6429$~km~s$^{-1}$, its radial velocity is more than $3\sigma$ away from the mean radial velocity defined by our dwarf galaxy sample, but it is still in the range of radial velocities covered by the other cluster galaxies in both the dwarf galaxy sample and the CZ03 sample (cf. Fig. \ref{fig:vrad}). However, with $R=5.65$ the significance of the cross-correlation match is one of the lowest in the VIMOS-data. Because the slit was not completely centred on the object (due to the mask misalignment in the second VIMOS-quadrant), the S/N ratio of the spectrum was relatively low and the CCF peak was less well visible. Hence, the cluster membership has to be reconfirmed with a higher S/N spectrum.

Note that, based on a purely photometric study, \citet{Wehner2007} have recently reported on the discovery of 29 UCD candidates around NGC 3311. They have luminosities of $-12.3<M_{g'}<-10.7$ mag. They build up the extension of the red GC population towards very high luminosities. Their cluster membership has to be confirmed by radial velocity measurements.

\subsection{The colour--magnitude relation of early-type galaxies}
\label{sec:hydracmd}
\begin{figure}
	\resizebox{\hsize}{!}{\includegraphics{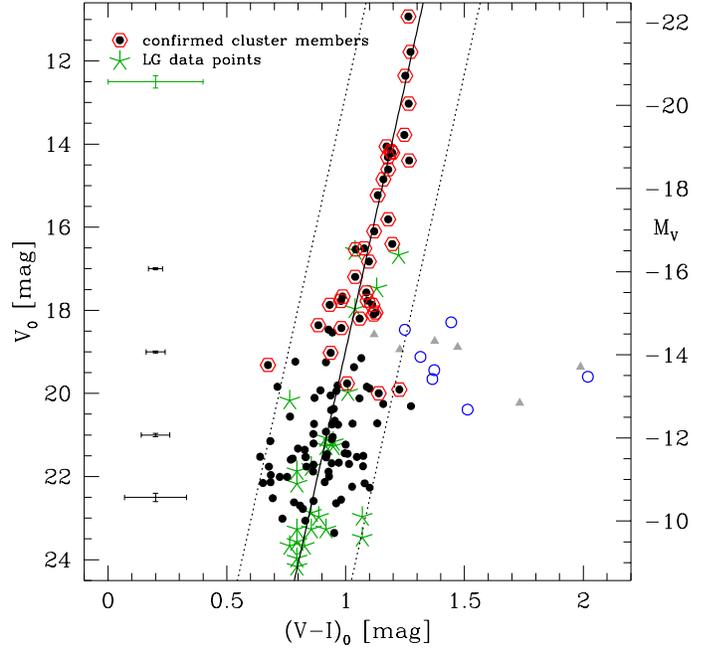}}
	\caption{Colour--magnitude diagram of early-type galaxies in the Hydra\,I cluster in comparison with Local Group dEs and dSphs. Black filled circles are probable cluster galaxies, selected by their morphology. Red open hexagons mark cluster members, confirmed by radial velocity measurements. Blue open circles are presumable background galaxies (see text for futher explanations). Grey triangles are confirmed background elliptical galaxies with $R^{1/4}$ surface brightness profiles. Typical errorbars are indicated. The solid line is a linear fit to dwarf galaxies with $M_V>-17$ mag (Eq. (\ref{eq:cmrhydra})). Dotted lines are the $2\sigma$ deviations from the fit. Green asterisks represent the Local Group dEs and dSphs (data from \citealt{Grebel2003}) projected to the Hydra\,I distance. Mean errors for the LG dwarfs are indicated in the upper left corner.}
	\label{fig:cmdhydra}
\end{figure}

A well-defined CMR for cluster galaxies brighter than $V\sim 20$ mag was already shown in the analysis of the spectroscopic data (Sect. \ref{sec:specresult}). Fig. \ref{fig:cmdhydra} shows the colour--magnitude diagram (CMD) of all early-type galaxies in our photometric sample (E and S0 as well as dE/dSph galaxies), as listed in Table \ref{tab:hydrasample}. Clearly, the CMR stretches across the entire magnitude range of $11<V_0<23$ mag, from the brightest giant elliptical galaxies all the way down to the regime of dwarf galaxies. The more luminous galaxies are on average redder than galaxies of lower luminosity. Adopting a distance modulus of $(m-M)=33.07$ mag \citep{Mieske2005}, a linear fit to \emph{all} data points leads to $(V-I)_0 = -0.040\cdot M_{V,0}+0.44$ with a rms of 0.12. The larger scatter at the faint magnitudes is consistent with the larger error in $(V-I)$. The magnitude limit at which faint dwarf galaxies could still be identified is $M_V\sim -10$ mag, comparable to the Local Group dSph Sculptor \citep{Grebel2003}. Due to the detection and resolution limit of our data, we cannot test for the existence of even fainter galaxies in Hydra\,I.

Colour--magnitude relations for dwarf galaxies have been observed in a number of other nearby galaxy clusters, such as Coma, Virgo, Perseus and Fornax \citep[e.g.][]{Secker1997, Conselice2003, Hilker2003, vanZee2004, Adami2006, Mieske2007, Lisker2008}. The CMR for early-type dwarf galaxies in Fornax is given by $(V-I)_{\mathrm{Fornax}} = -0.033\cdot M_{V,0}+0.52$ in the magnitude range $-17<M_V<-9$ mag \citep{Mieske2007}. Restricting the fit for the Hydra\,I sample to $M_V>-17$ mag leads to:
\begin{equation}
\label{eq:cmrhydra}
 (V-I)_0 = -0.039\cdot M_{V,0}+0.45
\end{equation}
with a rms of 0.12, as indicated by the solid line in Fig. \ref{fig:cmdhydra}. This is in good agreement with the relation found in Fornax. Moreover, it is almost indistinguishable from the CMR defined by the whole Hydra\,I sample. We note that the adopted distance modulus of $(m-M)=33.07$ mag is a comparatively low value. Other recent publications give a larger distance to Hydra\,I with a mean distance modulus of $(m-M)=33.37$ mag \citep[see also discussion in][]{Mieske2005}, but using the higher value shifts the CMR only marginally (by 0.01 mag) towards the blue.

In addition, Eq. (\ref{eq:cmrhydra}) is compared with the CMR of LG dwarf ellipticals and dwarf spheroidals. Homogeneous $(V-I)$ colours for LG dwarfs do not exist \citep[see][]{Mateo1998}, but they can be calculated from their average iron abundances. Assuming single stellar populations, \citet{Hilker2003} transform the average iron abundances from \citet{Grebel2003} to $(V-I)$ colours using Eq. (4) given in \citet{Kissler-Patig1998}. It turns out that the colours, estimated in this way, match remarkably well the CMR found for the Hydra\,I dwarf galaxies. A linear fit to the LG data gives $(V-I)_{\mathrm{LG}} = -0.038\cdot M_V +0.48$ with a rms of 0.09, matching very well the Hydra\,I CMR.

However, one has to be aware of the uncertainties in this analysis. Equation (4) in \citet{Kissler-Patig1998} describes a linear relation between the average iron abundances and colours of globular clusters. The direct application to \emph{other} stellar systems like dwarf galaxies is, in the first instance, not evident. But the assumption of a single stellar population seems to be a good approximation in most cases, since nearby dwarf galaxies are mostly dominated by an old stellar population. Integrated $(V-I)$ colours are tabulated for eight LG dwarfs \citep{Mateo1998}. The comparison of the transformed colours with the measured colours from \citet{Mateo1998} shows only small discrepancies of about $0.1$ mag. The errors for the transformed colours, as indicated in Fig. \ref{fig:cmdhydra}, are of the order of $\Delta(V-I)=0.2$ mag.

\begin{figure}
	\resizebox{\hsize}{!}{\includegraphics{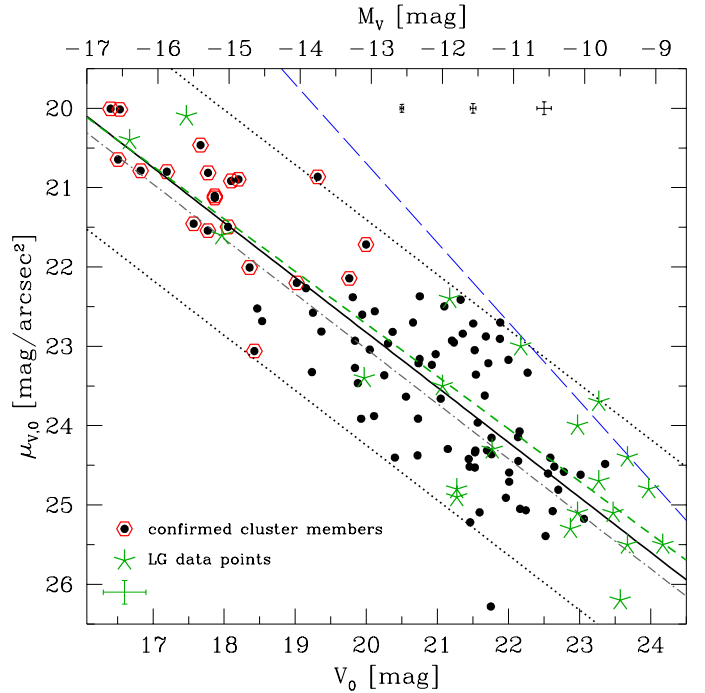}}
	\caption{Magnitude--surface brightness diagram of all early-type dwarf galaxies ($M_V>-17$ mag) in our sample in comparison with Local Group dEs and dSphs. The symbols are as in Fig. \ref{fig:cmdhydra}. Typical errorbars are indicated. The solid line is a linear fit to the data (Eq. (\ref{eq:murelhydra})). Dotted lines are the $2\sigma$ deviations from the fit. The dash-dotted line is a fit to the same data, but using a higher distance modulus (see text for details). Green asterisks represent Local Group dEs and dSphs (data from \citealt{Grebel2003}) projected to the Hydra\,I distance. The green dashed line is the LG magnitude--surface brightness relation. Mean errors for the LG dwarfs are indicated in the lower left corner. The blue long-dashed line indicates a scale length of $0.7\arcsec$ for an exponential profile, representing the resolution limit of our images.}
	\label{fig:muvhydra}
\end{figure}

A few remarks about the sample selection have to be made at this point. Object selection solely based on morphological classification can lead to the contamination of the sample with background galaxies that only resemble cluster dwarf elliptical galaxies. Bright background galaxies ($M_V\lesssim-15$ mag) could be excluded from our sample on the basis of radial velocity measurements. Towards the faintest magnitudes ($M_V\gtrsim-12$ mag), dwarf galaxies tend to be extended objects with very low surface brightnesses and less concentrated light profiles (typical of dSphs). Apparently small objects of the same apparent magnitude with a high central concentration of light or barely resolved objects, both being likely background galaxies, were therefore excluded (see Sect. \ref{sec:lfunction} for more details). In the intermediate magnitude range, the distinction between cluster dEs and background elliptical galaxies was more difficult. As indicated in Fig. \ref{fig:cmdhydra} by blue open circles, we found seven objects with $-14.8<M_V<-12.7$ mag, appearing very similar to confirmed cluster dEs in terms of their morphology. A first indication that they likely do not belong to the cluster is that they have significantly redder colours than other objects in the same magnitude range. Some of them are even too red to be a galaxy at $z\sim 0$. Moreover, they have surface brightness profiles that follow the de Vaucouleurs law ($R^{1/4}$ law), which is typical of giant elliptical galaxies. The comparison with spectroscopically confirmed background galaxies in the same magnitude range, which also exhibit $R^{1/4}$ surface brightness profiles, shows that they are the same group (see Fig. \ref{fig:cmdhydra}). Hence, it is reasonable to assume that the seven arguable objects only \emph{resemble} cluster dwarf galaxies, but they are in fact background giant elliptical galaxies.

\subsection{The magnitude--surface brightness relation}
\label{sec:magmurel}
\begin{figure}
	\resizebox{\hsize}{!}{\includegraphics{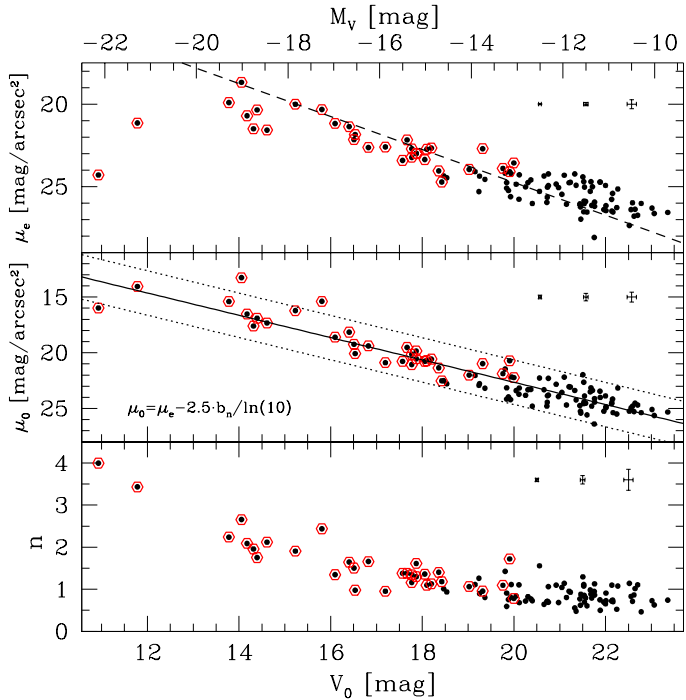}}
	\caption{Results of the S\'ersic fits: effective surface brightness $\mu_{\mathrm{e}}$ (top panel), central surface brightness $\mu_0$ (middle panel) and profile shape index $n$ (bottom panel) plotted vs. magnitude for all early-type galaxies in our sample. The dashed line in the upper panel indicates an effective radius of $4''$ (0.8 kpc at the cluster distance). The solid line in the middle panel is a linear fit to the data (Eq. (\ref{eq:sersicmurel})) with its $2\sigma$ deviations (dotted lines). Red open hexagons mark spectroscopically confirmed cluster members. Typical errorbars are indicated.}
	\label{fig:sersic}
\end{figure}

In Fig. \ref{fig:muvhydra}, the central surface brightness $\mu_{V,0}$, as estimated from an exponential law, is plotted vs. $M_V$ for all early-type galaxies fainter than $M_V=-17$ mag. A magnitude--surface brightness relation is visible in the sense that the central surface brightness increases with luminosity. A linear fit to the data yields
\begin{equation}
\label{eq:murelhydra}
 \mu_{V,0}= 0.69\cdot M_{V,0} + 31.88
\end{equation}
with a rms of 0.71. The relation is well-defined down to very low luminosities and surface brightnesses. The trend of the relation is similar to those found in other galaxy groups and clusters \citep[e.g.][]{Ferguson1988, Ulmer1996, Binggeli1998, Jerjen2000, Hilker2003, Adami2006, Mieske2007}. Local Group dEs and dSphs follow almost the same magnitude--surface brightness relation (data from \citealt{Grebel2003}). Using the larger distance modulus of $(m-M)=33.37$ mag changes the y-intercept of Eq. (\ref{eq:murelhydra}) by $+0.2$ mag (indicated by the dash-dotted line in Fig. \ref{fig:muvhydra}).

In order to include also the brighter ($M_V<-17$ mag) cluster early-type galaxies in the analysis, we additionally fitted \citet{Sersic1968} models to the galaxy surface brightness profiles. The effective surface brightness $\mu_{\mathrm{e}}$ against $M_V$ is shown in the upper panel of Fig. \ref{fig:sersic}. An increase of effective surface brightness with magnitude is visible between $-18 \lesssim M_V \lesssim -10$ mag. The effective radius stays virtually constant in this interval, as the (dashed) line, which represents a constant radius of $4''$ (0.8 kpc at the cluster distance), illustrates. This phenomenon has also been observed for dwarf galaxies in Coma, Virgo and most recently in the Antlia cluster \citep[][and references therein]{SmithCastelli2008}.

\begin{figure}
	\resizebox{\hsize}{!}{\includegraphics{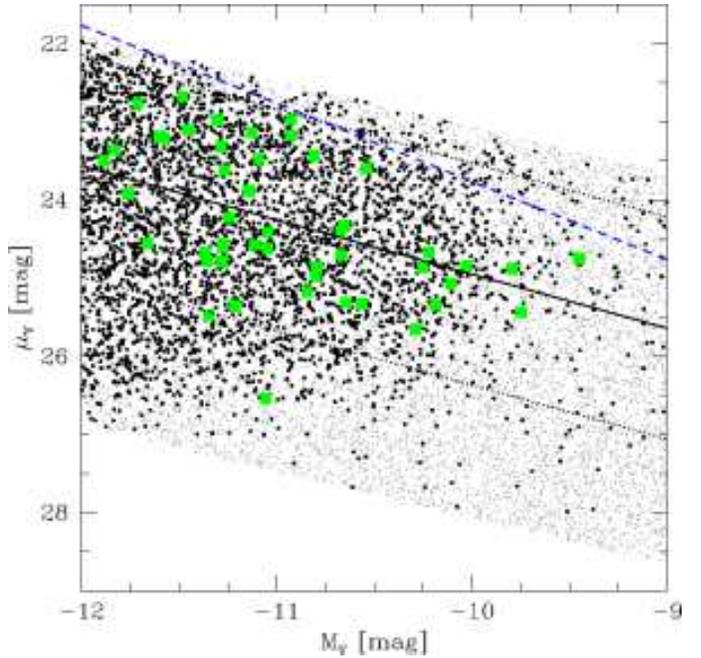}}
	\caption{Plot of the input-parameter range of the artificial dwarf galaxies (small grey dots). Black dots are the simulated galaxies recovered by SExtractor after applying several cuts (see text for further explanation). The objects from Fig. \ref{fig:muvhydra} are plotted as green squares. Equation (\ref{eq:murelhydra}) with its $2\sigma$ deviations is plotted as in Fig. \ref{fig:muvhydra}. The blue dashed line indicates a scale length of $0.7\arcsec$ for an exponential profile, representing the resolution limit of our images.}
	\label{fig:galsim}
\end{figure}

A different behaviour is observed at magnitudes brighter than $M_V \sim -18$ mag, in the sense that $\mu_{\mathrm{e}}$ levels off, with the exception of NGC 3311, the brightest galaxy in our sample. This has been reported by many authors in the past \citep[e.g.][]{Kormendy1985, Ferguson1988, Bender1992}, but according to \citet[][see their Fig. 12]{Graham2003} there is no dichotomy between dwarf galaxies and E/S0 galaxies when plotting the central surface brightness $\mu_0$ of a S\'{e}rsic model vs. the galaxy magnitude instead of $\mu_{\mathrm{e}}$ or $\langle\mu\rangle_{\mathrm{e}}$. $\mu_0$ is given by $\mu_0=\mu_{\mathrm{e}} - 2.5b_n/\ln(10)$, in which $b_n$ is approximated by $b_n = 1.9992n - 0.3271$ for $0.5<n<10$ \citep{Graham2005}. For our sample, $\mu_0$ vs. $M_V$ is shown in the middle panel of Fig. \ref{fig:sersic}. A continuous relation is visible for the low mass dwarf galaxies and the high mass Es and S0s ($M_V<-17$ mag). This continuity was also observed in the ACS Virgo and Fornax Cluster Surveys \citep{Ferrarese2006, Cote2006, Cote2007, Cote2008}. A linear fit to our data reveals a direct correlation between $\mu_0$ and $M_V$:
\begin{equation}
\label{eq:sersicmurel}
 \mu_{0}= 1.00\cdot M_{V,0} + 35.73
\end{equation}
with an rms of 1.00. Four of the brightest cluster galaxies (HCC-003, HCC-004, HCC-008, HCC-012) are excluded from the analysis, since their surface brightness profiles could not reasonably be fitted by a single S\'ersic profile, but rather showed two components (bulge + disk). These galaxies are morphologically classified as SAB(s)0, SB(rs)0, SB(s)0, and S(rs)0 respectively. However, they closely follow the cluster CMR (cf. Table \ref{tab:hydrasample} and Fig. \ref{fig:cmdhydra}).

The lower panel of Fig. \ref{fig:sersic} shows how the S\'{e}rsic profile shape index $n$ varies with magnitude. An increase of $n$ with increasing magnitude is visible for the magnitude range $-22\lesssim M_V\lesssim-14$ mag, whereas the data points scatters around a mean of $n=0.9$ for $M_V\gtrsim -14$ mag. A similar result was obtained for early-type galaxies in the Fornax cluster \citep{Infante2003}.

\subsection{The faint end of the galaxy luminosity function}
\label{sec:lfunction}
\begin{figure}
	\resizebox{\hsize}{!}{\includegraphics{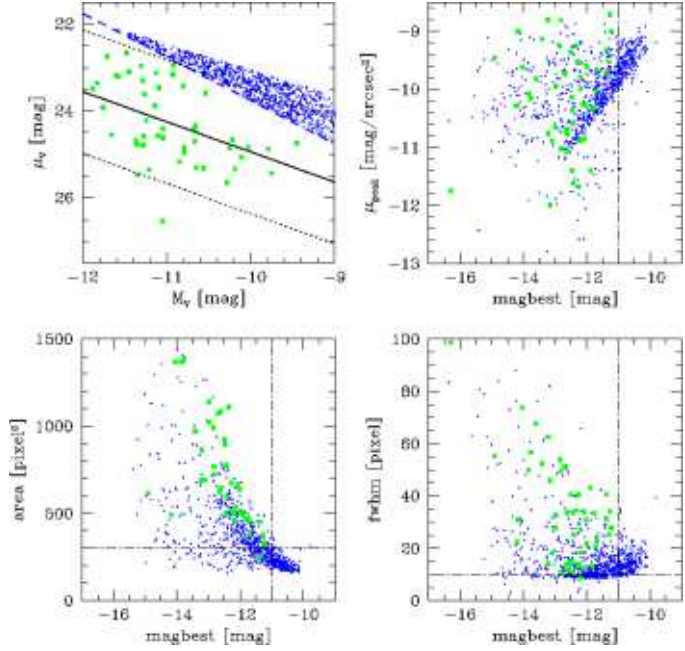}}
	\caption{SExtractor output-parameters of the artificial galaxies below the resolution limit (blue dots). The upper left panel shows the input-parameters $M_V$ and $\mu_V$. The SExtractor output-parameter \texttt{magbest} is plotted against \texttt{mupeak} (upper right), \texttt{area} (lower left) and \texttt{fwhm} (lower right). Green squares are the objects from Fig. \ref{fig:muvhydra}. Dash-dotted lines indicate the global cuts on \texttt{magbest}, \texttt{fwhm} and \texttt{area}.}
	\label{fig:artgalanalysis}
\end{figure}

\begin{figure}
	\resizebox{\hsize}{!}{\includegraphics{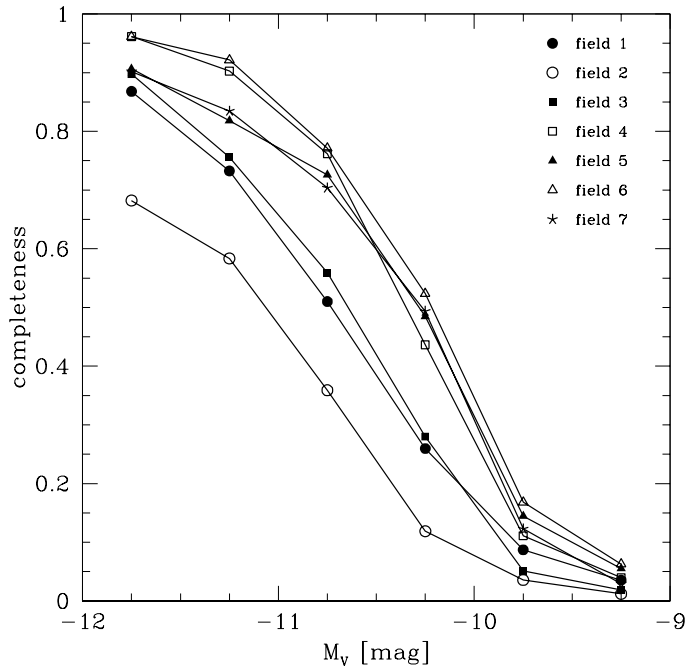}}
	\caption{Completeness as a function of magnitude for the seven observed fields. The low completeness in field 2 is due to the high geometrical incompleteness.}
	\label{fig:incompleteness}
\end{figure}

For the study of the faint end of the galaxy luminosity function, the number counts of dE/dSph candidates have to be completeness corrected. For this, we randomly distributed $10\,000$ simulated dwarf galaxies (in 500 runs) in each of the seven CCD fields, using a C++ code. The magnitudes and central surface brightnesses of the simulated galaxies were chosen such that they extended well beyond the observed parameter space at the faint limits. Exemplary for one field, Fig. \ref{fig:galsim} shows the input-parameter range of the simulated galaxies. SExtractor was then used to recover the artificial galaxies, and the SExtractor output-parameters, i.e. \texttt{magbest}, \texttt{mupeak}, \texttt{fwhm}, \texttt{area}, were compared with the parameters derived from the sample of actual cluster dwarf galaxies, as described in Sects. \ref{sec:magmurel} and \ref{sec:hydracmd}.

As already mentioned in Sect. \ref{sec:photobservation}, we applied several cuts to the SExtractor output-parameters with the aim of rejecting high surface brightness and barely resolved background objects. These objects were defined to be located above the (blue) dashed line in Fig. \ref{fig:galsim}, which represents the resolution limit of our images. Fig. \ref{fig:artgalanalysis} shows the SExtractor output-parameters of the artificial galaxies below this resolution limit. They define well localised areas in plots of \texttt{magbest} versus \texttt{mupeak}, \texttt{area} and \texttt{fwhm}. However, also some of the previously selected dwarf galaxy candidates scatter into the same areas. Hence, we rejected only those objects that \emph{simultaneously} occupied the locus of unresolved galaxies in all three parameters \texttt{mupeak}, \texttt{area} and \texttt{fwhm}.

In this way we miss only one of the previously selected galaxies but we reject the majority of objects below the resolution limit. Additionally, we applied global cuts at the lower limits of \texttt{magbest}, \texttt{fwhm} and \texttt{area} in order to reject very faint, almost unresolved background objects (see Fig. \ref{fig:artgalanalysis}). All artificial galaxies that were recovered after the application of the cuts are highlighted in Fig. \ref{fig:galsim}.

Without applying any cuts, SExtractor recovers 85--95\% of the artificial galaxies at $M_V\le-12$ mag, except for field 2 where the completeness is only 70\%, due to the large spiral galaxy in the field (cf. Figures \ref{fig:hydracluster} and \ref{fig:fields}). This reflects the geometrical incompleteness caused by blending. By applying the cuts in \texttt{magbest}, \texttt{mupeak}, \texttt{area} and \texttt{fwhm} we further reject about 15\% of the artificial galaxies at $M_V=-12$ mag. The fraction of visually classified galaxies with $M_V>-12$ mag that are excluded by applying the same cuts is 6 out of 43. This fraction is consistent with the fraction of excluded artificial galaxies. Since all visually selected galaxies are included into the LF, we scale the completeness values for $M_V>-12$ mag up by 15\%, so that they are consistent with the geometrical completeness at $M_V=-12$ mag. Fig. \ref{fig:incompleteness} shows the corrected completeness values in $0.5$ mag bins. The galaxy number counts are completeness corrected individually for each CCD field using these curves.

\begin{figure}
	\resizebox{\hsize}{!}{\includegraphics{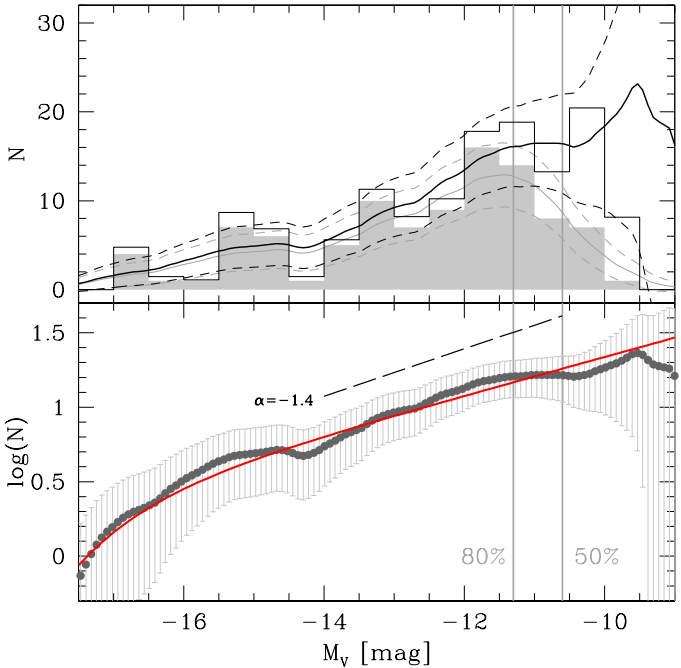}}
	\caption{Luminosity function of the Hydra\,I dwarf galaxies. The vertical lines mark the 80\% and 50\% completeness limits. Upper panel: the uncorrected galaxy number counts are displayed by the shaded histogram. The thin grey curve gives a binning independent representation for the counts (Epanechnikov kernel with 0.5 mag width). The completeness corrected number counts are given by the open histogram. The thick black curve represents the completeness corrected number counts with the $1\sigma$ uncertainties (dashed lines). Lower panel: completeness corrected number counts in logarithmic representation. The best fitting single Schechter function (solid red curve) is overlaid. A power-law slope of $\alpha=-1.4$ is indicated by the dashed line.}
	\label{fig:lumfunction}
\end{figure}

In Fig. \ref{fig:lumfunction} we show the resulting luminosity function of the Hydra\,I dwarfs in the magnitude range $-17.0<M_V<-9.5$ mag. By fitting a single \citet{Schechter1976} function to the number counts with a completeness larger than 50\%, we derive a faint-end slope of $\alpha=-1.13\pm 0.04$. The slope does not change if we include only data points with a completeness larger than 80\%. Alternatively, we fit a power-law model to the faint end of the LF. This results in $\alpha =-1.37 \pm 0.08$ for $-16.0<M_V<-10.6$ mag and $\alpha =-1.40 \pm 0.18$ for $-14.0<M_V<-10.6$ mag (indicated by the dashed line in Fig. \ref{fig:lumfunction}). Interestingly, the small dip in the galaxy LF at about $M_V=-14$ mag (although maybe the result of low number counts) appears near the luminosity where the separation of dEs and dSphs is defined \citep[e.g.][]{Grebel2001}. \citet{Hilker2003} reported on the same phenomenon in the Fornax cluster (see their Fig. 3).

\section{Summary and discussion}
\label{sec:discussion}
In this paper we presented a spectroscopic and photometric study of the early-type dwarf galaxy population in the Hydra\,I cluster. Two spectroscopic surveys were analysed, one executed with Magellan~I/LDSS2 at Las Campanas Observatory, the other with VLT/VIMOS at ESO/Paranal. The imaging data were obtained with VLT/FORS1. Seven fields were observed in Johnson $V$ and $I$ filters with a seeing between $0.5''$ and $0.7''$.

\subsection{Massive UCDs in Hydra\,I?}
From the two spectroscopic data sets we identify three compact cluster members being candidates for bright GCs and/or UCDs. Our rather incomplete VIMOS data set reveals the existence of one UCD candidate with $M_V=-13.26$ mag, corresponding to a mass of $\sim 10^8\ \mathrm{M_{\sun}}$. However, due to its comparatively low S/N spectrum its cluster membership has to be reconfirmed by follow-up observations.

This \emph{possible} detection is intriguing, since such sources are extremely rare in other environments. The complete spectroscopic surveys for Fornax and Virgo have revealed only one single such bright UCD in either cluster \citep{Jones2006}, with all other UCDs being at least one magnitude fainter. Does the possible detection of this bright UCD point to a large number of massive UCDs in Hydra\,I? The cluster centre is indeed dominated by the cD galaxy NGC 3311 with its very pronounced diffuse light component, which implies that the cluster environment in Hydra\,I could be more important in shaping the central region than in Fornax or Virgo. A large number of massive UCDs would therefore strengthen the idea of environment dependent UCD formation processes, such as the dynamical transformation of dwarf galaxies \citep{Bekki2003, Goerdt2008}. Further insight into the nature of the possible UCD candidate -- for example its mass-to-light ratio and size -- will require space based imaging and high resolution spectroscopy. With only one UCD candidate, it is not possible to make a meaningful statement about the abundance and properties of UCDs in the Hydra\,I cluster. For this, extended spectroscopic surveys are required.

\subsection{Global photometric properties of dwarf galaxies}
By radial velocity measurements we confirm the cluster membership of 29 galaxies, of which 5 are previously uncatalogued dwarf galaxies. The confirmed cluster galaxies define a tight colour--magnitude relation (Fig. \ref{fig:ldssprops}b). From the visual inspection of the images, we identify 82 additional cluster dE/dSph candidates. Our sample of $\simeq 100$ early-type cluster dwarf galaxies fainter than $M_V=-17$ mag follows a colour--magnitude relation being the extension of the CMR of the brighter cluster galaxies (see Fig. \ref{fig:cmdhydra}). This is consistent with earlier studies where early-type dwarf galaxies are reported to follow the CMR of giant early-type galaxies \citep[e.g.][]{Secker1997, Conselice2003, Hilker2003, LopezCruz2004, Adami2006}.

Our sample of early-type dwarf galaxies defines a magnitude--surface brightness relation in the sense that the central surface brightness (as estimated from an exponential law) increases with luminosity. Local Group dEs and dSphs follow reasonably well the same relation. By fitting S\'{e}rsic models to the galaxies surface brightness profiles, we find a common relation for dwarf galaxies and E/S0 galaxies in a $\mu_0$ -- $M_V$ diagram, consistent with the results of \citet{Graham2003}, \citet{Ferrarese2006} and \citet{Cote2006, Cote2007, Cote2008}. The slope of the relation is found to be equal to one. Moreover, in a $\mu_{\mathrm{e}}$ -- $M_V$ diagram, galaxies fainter than $M_V = -18$ mag define to a good approximation a relation of constant effective radius $R_{\mathrm{e}}\sim 0.8$ kpc (see Fig. \ref{fig:sersic}).

Our results, namely the \emph{common} relations for dwarf galaxies and giant elliptical galaxies in a colour -- magnitude and a  $\mu_0$ -- $M_V$ diagram over a wide range of magnitudes, suggest that dEs are not a separate class of objects but rather the low mass counterparts of massive early-type galaxies. The almost constant effective radius of galaxies with magnitudes $-18<M_V<-10$ mag might indicate the same.

We find a similarity of the Hydra\,I CMR to both the Fornax and the LG relation. Since the environments in Hydra\,I and Fornax are different from the Local Group in terms of mean density and strength of the gravitational potential \citep{Girardi1993, Girardi1998, Karachentsev2006}, this could imply that internal evolution is more important for the global photometric properties of dwarf galaxies than external influences due to different environments. The almost identical magnitude--surface brightness relations of the Hydra\,I and the LG dwarfs may support this hypothesis. \citet{SmithCastelli2008} argue in the same manner based on the results of their photometric study of the galaxy population in the Antlia cluster. Further deep photometric studies in various clusters of different mass and dynamical state have to be performed in order to investigate the influence of different environments on the photometric and structural properties of dwarf galaxies.

\subsection{The Hydra\,I LF}
We derive a very flat luminosity function ($\alpha \sim -1.1$, as derived from a Schechter fit) for the Hydra\,I cluster. Recently, \citet{Yamanoi2007} found $\alpha \sim -1.6$ for the cluster LF. Our observed fields overlap with their central observed region and we reach a similar limiting magnitude. The seeing of our images is slightly better. However, \citet{Yamanoi2007} do not fit Schechter functions to their data but use a power-law model to determine the parameter $\alpha$ at the faint end of the LF. In an analogous manner we derive a slope of $\alpha \sim -1.4$ from our data, being more consistent with the results of Yamanoi et al. Qualitatively, we find the same behaviour for the Fornax LF \citep{Hilker2003, Mieske2007}. When fitting a power-law to the faint end, we derive $\alpha \sim -1.3$ for $-14.0<M_V<-9.8$ mag, as opposed to $\alpha \sim -1.1$ when fitting a Schechter function. These findings can be interpreted in the way that the description by a power-law model seems to result in steeper slopes than the description by Schechter functions. The slope of the power-law model might be considered as an upper limit in this context.

We point out that the differences in $\alpha$ could also arise from the different methods used to construct the LF. The dwarf galaxy candidates in \citet{Yamanoi2007} were selected only by setting lower limits in FWHM to the SExtractor detections. Contaminating background galaxies were statistically subtracted. No curve-of-growth analysis or surface brightness profile fitting was performed to cross-check the SExtractor results. By means of the surface brightness profile and the colour we could exclude a number of objects from our sample (see Sect. \ref{sec:hydracmd} and Fig. \ref{fig:cmdhydra}). Our sample should therefore not be contaminated by many background galaxies that only resemble cluster dwarf galaxies.

Being consistent with our results, \citet{Trentham2002} measured the LF in different local environments (including the Virgo cluster) and found shallow logarithmic slopes of $\alpha \sim -1.2$. They applied selection criteria similar to ours (based on surface brightness and morphology). \citet{Trentham2002} stated that it is unlikely that they have missed a large number of dwarfs, since they were sensitive to very low surface brightness galaxies and the seeing was good enough to distinguish high surface brightness dwarfs from background galaxies. The same is true for our data. We could potentially have missed very compact cluster members that resemble objects like M32, but these objects are rare \citep{Drinkwater2000b, Mieske2005b, Chilingarian2007, Chilingarian2008} and do not significantly contribute to the LF.

Clearly, for the determination of the faint-end of the LF it is crucial which method is used to identify contaminating background galaxies. Different methods in constructing the LF can lead to different results. One example for this is the Fornax cluster. Applying morphological selection criteria, \citet{Hilker2003} found a flat LF, whereas \citet{Kambas2000} reported on a very steep LF ($\alpha \simeq -2$) for the same cluster. However, by comparing the data sets \citet{Hilker2003} could show that most of Kambas' dwarf galaxy candidates were non-members of the Fornax cluster.

\vspace{0.25cm}
\noindent
We conclude that at least for nearby galaxy clusters that are close enough for large telescopes like the VLT to resolve even faint dwarf spheroidal galaxies under good seeing conditions, the membership assignment by means of morphology and surface brightness seems to be an appropriate way to construct the galaxy luminosity function and constrain photometric scaling relations.

\bibliographystyle{aa}
\bibliography{hydradwarfs}

\Online

\begin{appendix}

\section{Tables}
Table \ref{tab:calcoeff} gives the photometric calibration coefficients for the 7 observed fields as indicated in Fig. \ref{fig:fields}. Zero points (ZP), extinction coefficients $k$ and colour terms (CT) are given for the two filters $V$ and $I$.

Table \ref{tab:hydrasample} lists the photometric parameters of our sample of 111 early-type galaxies in the Hydra\,I cluster. The table is ordered by increasing apparent magnitude. The first column gives the object ID, in which HCC stands for Hydra\,I Cluster Catalogue. Right ascension and declination (J2000.0) are given in columns two and three. The fourth and fifth column list the extinction corrected magnitude $V_0$ and the colour $(V-I)_0$. In columns six and seven, the central surface brightness $\mu_{V,0}$ and the scale length $h_R$ of an exponential fit to the surface brightness profile are listed. For objects with $V_0\leq16.1$ mag, $\mu_{V,0}$ and $h_R$ are not given, since the surface brightness profile is not well described by an exponential law. Columns eight, nine and ten give the effective surface brightness $\mu_{\mathrm{e}}$, the effective radius $R_{\mathrm{e}}$ and the profile shape index $n$, as obtained from a S\'ersic fit. The physical scale is 0.2 kpc/arcsec at the assumed distance modulus of $(m-M)=33.07$ mag \citep{Mieske2005}. The last column gives the radial velocities derived in our study as well as those from CZ03.

\begin{table}[hb!]
	\caption{Photometric calibration coefficients.}
	\label{tab:calcoeff}
	\centering	
		\begin{tabular}{l r r r r r r}
		\hline\hline
		Field & $\mathrm{ZP}_V$ & $\mathrm{ZP}_I$ & $k_V$ & $k_I$ & $\mathrm{CT}_V$ & $\mathrm{CT}_I$ \\
		\hline
		1 & 27.477 & 26.629 & -0.160 & -0.090 & 0.04 & -0.04 \\
		2 & 27.529 & 26.643 & -0.160 & -0.090 & 0.04 & -0.04 \\
		3 & 27.529 & 26.643 & -0.160 & -0.090 & 0.04 & -0.04 \\
		4 & 27.529 & 26.643 & -0.160 & -0.090 & 0.04 & -0.04 \\
		5 & 27.532 & 26.665 & -0.160 & -0.090 & 0.04 & -0.04 \\
		6 & 27.532 & 26.665 & -0.160 & -0.090 & 0.04 & -0.04 \\
		7 & 27.532 & 26.679 & -0.160 & -0.090 & 0.04 & -0.04 \\
		\hline
		\end{tabular}
\end{table}

\clearpage
\onecolumn

{\tiny
\longtab{2}{
\begin{longtable}{lccccc@{\hspace{1mm}}c@{\hspace{2mm}}c@{\hspace{1mm}}c@{\hspace{3mm}}c@{\hspace{3mm}}l}
\caption{\label{tab:hydrasample} The Hydra\,I Cluster Catalogue (HCC).}\\
\hline\hline
ID & $\alpha$(2000.0) & $\delta$(2000.0) & $V_0$ & $(V-I)_0$ & $\mu_{V,0}$ & $h_R$ & $\mu_{\mathrm{e}}$ & $R_{\mathrm{e}}$ & $n$ & $v_{\mathrm{rad}}$ \\
~ & [h:m:s] & [$^\circ$:$\arcmin$:$\arcsec$] & [mag] & [mag] & [mag arcsec$^{-2}$] & [arcsec] & [mag arcsec$^{-2}$] & [arcsec] & ~ & [km s$^{-1}$] \\
\hline
\endfirsthead
\caption{continued.}\\
\hline\hline
ID & $\alpha$(2000.0) & $\delta$(2000.0) & $V_0$ & $(V-I)_0$ & $\mu_{V,0}$ & $h_R$ & $\mu_{\mathrm{e}}$ & $R_{\mathrm{e}}$ & $n$ & $v_{\mathrm{rad}}$ \\
~ & [h:m:s] & [$^\circ$:$\arcmin$:$\arcsec$] & [mag] & [mag] & [mag arcsec$^{-2}$] & [arcsec] & [mag arcsec$^{-2}$] & [arcsec] & ~ & [km s$^{-1}$] \\
\hline
\endhead
\hline
\endfoot

HCC-001$^c$ & 10:36:42.8 & -27:31:42.0 & $10.93 \pm 0.01$ & $1.26 \pm 0.02$ &  &  & $24.30 \pm 0.01$ & $150.25 \pm 0.34$ & 4.00 & $3818 \pm 45$ \\ 
HCC-002$^d$ & 10:36:35.7 & -27:31:06.4 & $11.78 \pm 0.02$ & $1.27 \pm 0.04$ &  &  & $21.15 \pm 0.01$ & $19.35 \pm 0.01$ & 3.43 & $4099 \pm 27$ \\ 
HCC-003$^{b,e}$ & 10:36:22.4 & -27:26:17.3 & $12.36 \pm 0.01$ & $1.25 \pm 0.01$ &  &  &  &  &  & $3614 \pm 29$\\ 
HCC-004$^{b,f}$ & 10:37:37.3 & -27:35:39.4 & $13.03 \pm 0.01$ & $1.26 \pm 0.01$ &  &  &  &  &  & $3976 \pm 14$\\ 
HCC-005 & 10:36:27.7 & -27:19:10.8 & $13.78 \pm 0.01$ & $1.25 \pm 0.01$ &  &  & $19.90 \pm 0.01$ & $5.84 \pm 0.01$ & 2.24 & $3376 \pm 20$\\ 
HCC-006 & 10:36:44.9 & -27:28:10.1 & $14.06 \pm 0.01$ & $1.17 \pm 0.01$ &  &  & $18.67 \pm 0.02$ & $2.23 \pm 0.01$ & 2.66 & $2735\pm80^a$\\ 
HCC-007 & 10:36:41.2 & -27:33:39.6 & $14.18 \pm 0.01$ & $1.19 \pm 0.01$ &  &  & $20.70 \pm 0.01$ & $7.36 \pm 0.01$ & 2.09 & $4830 \pm 13$\\ 
HCC-008$^b$ & 10:37:20.1 & -27:33:35.6 & $14.21 \pm 0.01$ & $1.20 \pm 0.01$ &  &  &  &  &  & $4069 \pm 14$\\ 
HCC-009 & 10:36:29.0 & -27:29:02.2 & $14.32 \pm 0.01$ & $1.18 \pm 0.01$ &  &  & $21.49 \pm 0.01$ & $9.00 \pm 0.02$ & 1.96 & $4774\pm80^a$\\ 
HCC-010 & 10:36:23.0 & -27:21:16.8 & $14.40 \pm 0.01$ & $1.27 \pm 0.01$ &  &  & $20.35 \pm 0.01$ & $4.88 \pm 0.01$ & 1.75 & $3059 \pm 18$\\ 
HCC-011 & 10:36:24.8 & -27:34:54.8 & $14.61 \pm 0.01$ & $1.18 \pm 0.01$ &  &  & $21.57 \pm 0.00$ & $7.54 \pm 0.02$ & 2.12 & $2895 \pm 20$\\ 
HCC-012$^b$ & 10:36:35.0 & -27:28:45.4 & $14.85 \pm 0.01$ & $1.16 \pm 0.01$ &  &  &  &  &  & $4708 \pm 17$\\ 
HCC-013 & 10:36:35.5 & -27:28:13.6 & $15.23 \pm 0.01$ & $1.13 \pm 0.01$ &  &  & $20.01 \pm 0.01$ & $2.63 \pm 0.01$ & 1.91 & $3199 \pm 13$\\ 
HCC-014 & 10:36:49.1 & -27:23:20.0 & $15.81 \pm 0.01$ & $1.18 \pm 0.01$ &  &  & $20.32 \pm 0.01$ & $2.17 \pm 0.01$ & 2.44 & $4474\pm80^a$\\ 
HCC-015 & 10:37:38.1 & -27:35:58.8 & $16.10 \pm 0.01$ & $1.12 \pm 0.01$ &  &  & $21.17 \pm 0.00$ & $3.05 \pm 0.01$ & 1.35 & $3995 \pm 14$\\ 
HCC-016 & 10:36:56.9 & -27:34:03.5 & $16.40 \pm 0.01$ & $1.20 \pm 0.02$ & $20.01 \pm 0.03$ & $2.10 \pm 0.02$ & $21.36 \pm 0.01$ & $2.91 \pm 0.01$ & 1.64 & $4468 \pm 24$\\ 
HCC-017 & 10:36:42.7 & -27:35:08.7 & $16.51 \pm 0.01$ & $1.08 \pm 0.01$ & $20.64 \pm 0.01$ & $3.23 \pm 0.02$ & $22.15 \pm 0.01$ & $4.74 \pm 0.02$ & 1.50 & $3572 \pm 29$\\ 
HCC-018 & 10:37:47.0 & -27:34:21.8 & $16.53 \pm 0.01$ & $1.04 \pm 0.01$ & $20.01 \pm 0.01$ & $2.62 \pm 0.02$ & $21.84 \pm 0.01$ & $4.39 \pm 0.01$ & 0.98 & $4861 \pm 53$\\ 
HCC-019 & 10:36:52.6 & -27:32:14.7 & $16.82 \pm 0.02$ & $1.10 \pm 0.03$ & $20.79 \pm 0.02$ & $2.53 \pm 0.02$ & $22.63 \pm 0.02$ & $4.24 \pm 0.04$ & 1.66 & $5735 \pm 55$\\ 
HCC-020 & 10:36:26.9 & -27:23:25.6 & $17.19 \pm 0.01$ & $1.04 \pm 0.02$ & $20.80 \pm 0.01$ & $2.82 \pm 0.01$ & $22.60 \pm 0.01$ & $4.71 \pm 0.02$ & 0.95 & $4223 \pm 49$\\ 
HCC-021 & 10:37:17.3 & -27:35:33.7 & $17.57 \pm 0.03$ & $1.09 \pm 0.04$ & $21.45 \pm 0.02$ & $2.66 \pm 0.03$ & $23.41 \pm 0.02$ & $4.67 \pm 0.06$ & 1.38 & $3700 \pm 41$\\ 
HCC-022 & 10:36:40.4 & -27:32:56.5 & $17.67 \pm 0.03$ & $0.99 \pm 0.04$ & $20.46 \pm 0.02$ & $1.56 \pm 0.02$ & $22.17 \pm 0.01$ & $2.46 \pm 0.01$ & 1.38 & $4605 \pm 37$\\ 
HCC-023 & 10:36:48.9 & -27:30:01.9 & $17.77 \pm 0.04$ & $1.09 \pm 0.07$ & $21.54 \pm 0.07$ & $2.52 \pm 0.07$ & $23.23 \pm 0.02$ & $3.99 \pm 0.04$ & 1.16 & $4479 \pm 44$\\ 
HCC-024 & 10:36:50.2 & -27:30:47.7 & $17.77 \pm 0.02$ & $0.98 \pm 0.03$ & $20.81 \pm 0.02$ & $1.71 \pm 0.02$ & $22.71 \pm 0.02$ & $2.94 \pm 0.03$ & 1.35 & $5270 \pm 32$\\ 
HCC-025 & 10:37:13.7 & -27:30:25.2 & $17.87 \pm 0.05$ & $1.11 \pm 0.07$ & $21.10 \pm 0.03$ & $1.79 \pm 0.02$ & $22.99 \pm 0.03$ & $3.04 \pm 0.05$ & 1.61 & $2879 \pm 37$\\ 
HCC-026 & 10:36:46.0 & -27:31:25.1 & $17.87 \pm 0.04$ & $0.93 \pm 0.05$ & $21.13 \pm 0.03$ & $2.02 \pm 0.03$ & $23.00 \pm 0.04$ & $3.43 \pm 0.06$ & 1.30 & $4225\pm159^a$\\ 
HCC-027 & 10:36:45.7 & -27:30:32.1 & $18.05 \pm 0.06$ & $1.13 \pm 0.09$ & $21.49 \pm 0.04$ & $2.03 \pm 0.04$ & $23.36 \pm 0.02$ & $3.42 \pm 0.04$ & 1.36 & $5251\pm89^a$\\ 
HCC-028 & 10:36:43.1 & -27:25:30.1 & $18.10 \pm 0.01$ & $1.12 \pm 0.01$ & $20.92 \pm 0.01$ & $1.59 \pm 0.01$ & $22.73 \pm 0.01$ & $2.65 \pm 0.01$ & 1.10 & $3937 \pm 38$\\ 
HCC-029 & 10:37:00.2 & -27:29:53.4 & $18.20 \pm 0.01$ & $1.06 \pm 0.01$ & $20.90 \pm 0.02$ & $1.72 \pm 0.01$ & $22.66 \pm 0.01$ & $2.80 \pm 0.01$ & 1.13 & $3275 \pm 47$\\ 
HCC-030 & 10:36:50.8 & -27:23:02.3 & $18.36 \pm 0.01$ & $0.88 \pm 0.01$ & $22.01 \pm 0.02$ & $3.00 \pm 0.04$ & $24.05 \pm 0.01$ & $5.53 \pm 0.02$ & 1.40 & $4306\pm134^a$\\ 
HCC-031 & 10:38:04.6 & -27:32:44.9 & $18.42 \pm 0.01$ & $0.98 \pm 0.02$ & $23.06 \pm 0.01$ & $4.66 \pm 0.03$ & $24.73 \pm 0.01$ & $7.22 \pm 0.05$ & 1.18 & $2418\pm256^a$\\ 
HCC-032 & 10:36:26.0 & -27:35:51.6 & $18.47 \pm 0.01$ & $0.93 \pm 0.01$ & $22.52 \pm 0.01$ & $3.15 \pm 0.01$ & $24.35 \pm 0.01$ & $5.30 \pm 0.02$ & 1.01 & \\ 
HCC-033 & 10:37:48.9 & -27:33:03.0 & $18.53 \pm 0.02$ & $0.94 \pm 0.02$ & $22.68 \pm 0.01$ & $3.04 \pm 0.01$ & $24.47 \pm 0.01$ & $5.01 \pm 0.02$ & 0.94 & \\ 
HCC-034 & 10:36:58.6 & -27:34:26.3 & $19.02 \pm 0.09$ & $0.94 \pm 0.11$ & $22.20 \pm 0.07$ & $1.93 \pm 0.07$ & $23.96 \pm 0.10$ & $3.05 \pm 0.18$ & 1.06 & $4962 \pm 79$\\ 
HCC-035 & 10:36:32.5 & -27:32:23.3 & $19.15 \pm 0.02$ & $1.07 \pm 0.03$ & $22.27 \pm 0.02$ & $1.78 \pm 0.02$ & $24.10 \pm 0.01$ & $2.98 \pm 0.01$ & 1.11 & \\ 
HCC-036 & 10:36:41.3 & -27:23:36.7 & $19.23 \pm 0.02$ & $0.79 \pm 0.02$ & $23.32 \pm 0.02$ & $3.43 \pm 0.04$ & $25.31 \pm 0.01$ & $6.19 \pm 0.04$ & 1.26 & \\ 
HCC-037 & 10:36:43.1 & -27:35:43.5 & $19.25 \pm 0.01$ & $0.92 \pm 0.02$ & $22.58 \pm 0.02$ & $2.09 \pm 0.02$ & $24.37 \pm 0.01$ & $3.49 \pm 0.01$ & 0.90 & \\ 
HCC-038 & 10:36:28.7 & -27:14:36.4 & $19.32 \pm 0.01$ & $0.67 \pm 0.01$ & $20.86 \pm 0.01$ & $0.91 \pm 0.01$ & $22.70 \pm 0.01$ & $1.54 \pm 0.01$ & 0.96 & $3989\pm80^a$\\ 
HCC-039 & 10:36:38.8 & -27:25:07.9 & $19.37 \pm 0.02$ & $1.03 \pm 0.04$ & $22.81 \pm 0.02$ & $2.28 \pm 0.02$ & $24.57 \pm 0.01$ & $3.72 \pm 0.01$ & 0.80 & \\ 
HCC-040 & 10:37:39.6 & -27:35:32.7 & $19.76 \pm 0.02$ & $1.01 \pm 0.04$ & $22.14 \pm 0.03$ & $1.37 \pm 0.02$ & $23.90 \pm 0.01$ & $2.23 \pm 0.01$ & 1.10 & $2758 \pm 75$\\ 
HCC-041 & 10:36:43.7 & -27:17:29.9 & $19.81 \pm 0.01$ & $0.97 \pm 0.02$ & $22.38 \pm 0.02$ & $1.75 \pm 0.02$ & $24.19 \pm 0.01$ & $2.89 \pm 0.02$ & 1.42 & \\ 
HCC-042 & 10:36:35.9 & -27:19:36.2 & $19.84 \pm 0.01$ & $0.71 \pm 0.01$ & $23.27 \pm 0.03$ & $2.71 \pm 0.04$ & $25.10 \pm 0.01$ & $4.62 \pm 0.02$ & 0.59 & \\ 
HCC-043 & 10:36:40.6 & -27:25:19.1 & $19.84 \pm 0.03$ & $1.09 \pm 0.05$ & $22.93 \pm 0.02$ & $1.94 \pm 0.02$ & $24.76 \pm 0.01$ & $3.27 \pm 0.02$ & 0.91 & \\ 
HCC-044 & 10:36:45.1 & -27:14:29.4 & $19.88 \pm 0.03$ & $1.10 \pm 0.04$ & $23.47 \pm 0.03$ & $2.27 \pm 0.03$ & $25.24 \pm 0.02$ & $3.77 \pm 0.03$ & 0.76 & \\ 
HCC-045 & 10:36:44.4 & -27:31:22.8 & $19.91 \pm 0.12$ & $1.23 \pm 0.16$ & $24.66 \pm 0.17$ & $2.67 \pm 0.24$ & $24.10 \pm 0.02$ & $2.22 \pm 0.03$ & 1.72 & $4252 \pm 60$\\ 
HCC-046 & 10:37:23.1 & -27:35:56.7 & $19.93 \pm 0.02$ & $0.89 \pm 0.03$ & $23.91 \pm 0.02$ & $3.51 \pm 0.06$ & $25.62 \pm 0.03$ & $5.66 \pm 0.10$ & 0.81 & \\ 
HCC-047 & 10:36:33.8 & -27:27:41.1 & $19.94 \pm 0.03$ & $0.96 \pm 0.04$ & $22.60 \pm 0.04$ & $1.46 \pm 0.02$ & $24.23 \pm 0.01$ & $2.24 \pm 0.01$ & 1.11 & \\ 
HCC-048 & 10:37:11.3 & -27:31:30.3 & $20.01 \pm 0.02$ & $1.14 \pm 0.03$ & $21.72 \pm 0.03$ & $1.09 \pm 0.01$ & $23.57 \pm 0.01$ & $1.86 \pm 0.01$ & 0.78 & $2876 \pm 38$\\ 
HCC-049 & 10:37:56.0 & -27:32:41.9 & $20.05 \pm 0.02$ & $0.94 \pm 0.03$ & $23.04 \pm 0.02$ & $1.84 \pm 0.02$ & $24.86 \pm 0.02$ & $3.12 \pm 0.02$ & 0.84 & \\ 
HCC-050 & 10:37:39.1 & -27:33:42.3 & $20.11 \pm 0.06$ & $0.87 \pm 0.07$ & $23.88 \pm 0.02$ & $2.65 \pm 0.03$ & $25.77 \pm 0.05$ & $4.62 \pm 0.14$ & 1.11 & \\ 
HCC-051 & 10:38:04.7 & -27:32:22.5 & $20.12 \pm 0.01$ & $1.06 \pm 0.02$ & $22.56 \pm 0.03$ & $1.97 \pm 0.03$ & $24.29 \pm 0.01$ & $3.21 \pm 0.02$ & 0.68 & \\ 
HCC-052 & 10:36:55.4 & -27:35:55.1 & $20.25 \pm 0.01$ & $1.16 \pm 0.01$ & $23.36 \pm 0.03$ & $2.22 \pm 0.05$ & $25.02 \pm 0.03$ & $3.45 \pm 0.05$ & 0.72 & \\ 
HCC-053 & 10:37:03.7 & -27:35:47.7 & $20.30 \pm 0.02$ & $1.27 \pm 0.03$ & $22.96 \pm 0.04$ & $1.76 \pm 0.04$ & $24.76 \pm 0.02$ & $2.93 \pm 0.03$ & 0.82 & \\ 
HCC-054 & 10:36:44.9 & -27:23:01.4 & $20.37 \pm 0.01$ & $0.95 \pm 0.02$ & $22.82 \pm 0.02$ & $1.76 \pm 0.02$ & $24.59 \pm 0.01$ & $2.92 \pm 0.01$ & 0.77 & \\ 
HCC-055 & 10:36:37.2 & -27:22:54.7 & $20.40 \pm 0.07$ & $0.94 \pm 0.10$ & $24.40 \pm 0.03$ & $2.88 \pm 0.06$ & $26.03 \pm 0.04$ & $4.42 \pm 0.10$ & 0.73 & \\ 
HCC-056 & 10:37:51.2 & -27:32:39.4 & $20.56 \pm 0.05$ & $0.77 \pm 0.05$ & $23.63 \pm 0.09$ & $1.78 \pm 0.08$ & $25.29 \pm 0.03$ & $2.74 \pm 0.05$ & 1.56 & \\ 
HCC-057 & 10:36:48.2 & -27:28:52.2 & $20.66 \pm 0.01$ & $0.95 \pm 0.04$ & $22.70 \pm 0.06$ & $1.33 \pm 0.04$ & $24.36 \pm 0.02$ & $2.13 \pm 0.02$ & 0.64 & \\ 
HCC-058 & 10:36:41.5 & -27:16:37.3 & $20.72 \pm 0.08$ & $1.13 \pm 0.13$ & $24.38 \pm 0.03$ & $2.86 \pm 0.06$ & $25.96 \pm 0.04$ & $4.24 \pm 0.10$ & 0.70 & \\ 
HCC-059 & 10:37:12.5 & -27:29:52.3 & $20.73 \pm 0.05$ & $1.03 \pm 0.07$ & $23.91 \pm 0.04$ & $1.93 \pm 0.04$ & $25.65 \pm 0.04$ & $3.14 \pm 0.06$ & 0.71 & \\ 
HCC-060 & 10:38:03.0 & -27:35:46.9 & $20.73 \pm 0.03$ & $0.87 \pm 0.03$ & $23.21 \pm 0.03$ & $1.62 \pm 0.03$ & $25.04 \pm 0.02$ & $2.75 \pm 0.03$ & 0.71 & \\ 
HCC-061 & 10:36:29.3 & -27:27:23.7 & $20.75 \pm 0.04$ & $0.94 \pm 0.07$ & $23.16 \pm 0.03$ & $1.42 \pm 0.02$ & $24.94 \pm 0.02$ & $2.35 \pm 0.02$ & 0.69 & \\ 
HCC-062 & 10:36:31.0 & -27:17:58.5 & $20.75 \pm 0.01$ & $0.97 \pm 0.01$ & $22.37 \pm 0.03$ & $1.11 \pm 0.02$ & $24.23 \pm 0.02$ & $1.90 \pm 0.01$ & 1.06 & \\ 
HCC-063 & 10:38:06.3 & -27:35:25.5 & $20.92 \pm 0.04$ & $0.92 \pm 0.05$ & $23.23 \pm 0.05$ & $1.42 \pm 0.03$ & $24.85 \pm 0.02$ & $2.17 \pm 0.26$ & 1.10 & \\ 
HCC-064 & 10:36:29.8 & -27:33:27.0 & $20.97 \pm 0.05$ & $0.86 \pm 0.06$ & $23.10 \pm 0.05$ & $1.20 \pm 0.03$ & $25.06 \pm 0.03$ & $2.14 \pm 0.02$ & 0.69 & \\ 
HCC-065 & 10:36:43.7 & -27:32:57.7 & $21.05 \pm 0.03$ & $0.94 \pm 0.05$ & $23.66 \pm 0.04$ & $1.79 \pm 0.04$ & $25.47 \pm 0.05$ & $2.99 \pm 0.09$ & 1.04 & \\ 
HCC-066 & 10:37:41.6 & -27:32:21.6 & $21.10 \pm 0.02$ & $0.94 \pm 0.02$ & $22.50 \pm 0.02$ & $0.84 \pm 0.01$ & $24.31 \pm 0.03$ & $1.40 \pm 0.02$ & 1.14 & \\ 
HCC-067 & 10:36:35.8 & -27:15:51.5 & $21.15 \pm 0.08$ & $0.68 \pm 0.09$ & $24.29 \pm 0.04$ & $1.97 \pm 0.04$ & $26.08 \pm 0.07$ & $3.26 \pm 0.14$ & 0.80 & \\ 
HCC-068 & 10:36:25.9 & -27:17:23.6 & $21.20 \pm 0.01$ & $0.87 \pm 0.01$ & $22.93 \pm 0.06$ & $1.12 \pm 0.04$ & $24.77 \pm 0.04$ & $1.93 \pm 0.04$ & 0.95 & \\ 
HCC-069 & 10:36:23.8 & -27:13:19.2 & $21.23 \pm 0.03$ & $1.00 \pm 0.05$ & $22.95 \pm 0.01$ & $0.99 \pm 0.01$ & $24.74 \pm 0.02$ & $1.63 \pm 0.01$ & 0.93 & \\ 
HCC-070 & 10:36:47.3 & -27:24:13.2 & $21.33 \pm 0.02$ & $0.80 \pm 0.03$ & $22.41 \pm 0.09$ & $0.86 \pm 0.03$ & $24.24 \pm 0.02$ & $1.48 \pm 0.01$ & 0.56 & \\ 
HCC-071 & 10:36:45.6 & -27:34:57.4 & $21.36 \pm 0.05$ & $0.83 \pm 0.07$ & $22.84 \pm 0.12$ & $1.00 \pm 0.05$ & $24.92 \pm 0.02$ & $1.86 \pm 0.01$ & 0.48 & \\ 
HCC-072 & 10:36:48.3 & -27:26:46.7 & $21.44 \pm 0.07$ & $1.00 \pm 0.12$ & $24.42 \pm 0.04$ & $2.05 \pm 0.06$ & $26.06 \pm 0.13$ & $3.24 \pm 0.28$ & 0.77 & \\ 
HCC-073 & 10:36:31.0 & -27:15:15.0 & $21.45 \pm 0.06$ & $1.01 \pm 0.10$ & $24.52 \pm 0.05$ & $1.95 \pm 0.06$ & $26.24 \pm 0.07$ & $3.18 \pm 0.14$ & 0.62 & \\ 
HCC-074 & 10:36:24.0 & -27:27:46.9 & $21.46 \pm 0.08$ & $0.92 \pm 0.12$ & $25.22 \pm 0.05$ & $3.79 \pm 0.13$ & $26.98 \pm 0.20$ & $6.51 \pm 0.98$ & 0.88 & \\ 
HCC-075 & 10:36:22.9 & -27:35:41.5 & $21.50 \pm 0.04$ & $1.07 \pm 0.06$ & $22.71 \pm 0.05$ & $0.77 \pm 0.02$ & $24.42 \pm 0.05$ & $1.21 \pm 0.03$ & 1.30 & \\ 
HCC-076 & 10:36:26.5 & -27:32:16.0 & $21.52 \pm 0.05$ & $0.64 \pm 0.06$ & $23.05 \pm 0.06$ & $0.89 \pm 0.03$ & $24.70 \pm 0.04$ & $1.34 \pm 0.03$ & 1.11 & \\ 
HCC-077 & 10:36:21.3 & -27:31:22.0 & $21.52 \pm 0.10$ & $0.83 \pm 0.13$ & $24.53 \pm 0.06$ & $1.82 \pm 0.08$ & $26.53 \pm 0.37$ & $3.55 \pm 1.10$ & 1.08 & \\ 
HCC-078 & 10:36:40.8 & -27:19:39.0 & $21.53 \pm 0.05$ & $1.05 \pm 0.07$ & $24.34 \pm 0.08$ & $1.75 \pm 0.08$ & $25.88 \pm 0.10$ & $2.52 \pm 0.14$ & 0.84 & \\ 
HCC-079 & 10:36:26.6 & -27:27:12.6 & $21.53 \pm 0.06$ & $0.83 \pm 0.09$ & $24.31 \pm 0.06$ & $1.79 \pm 0.08$ & $26.04 \pm 0.07$ & $2.91 \pm 0.12$ & 0.86 & \\ 
HCC-080 & 10:36:24.6 & -27:11:43.9 & $21.54 \pm 0.04$ & $0.92 \pm 0.05$ & $23.36 \pm 0.04$ & $1.11 \pm 0.02$ & $25.03 \pm 0.02$ & $1.75 \pm 0.02$ & 0.69 & \\ 
HCC-081 & 10:36:26.5 & -27:32:41.6 & $21.57 \pm 0.06$ & $0.78 \pm 0.06$ & $23.96 \pm 0.08$ & $1.32 \pm 0.05$ & $25.90 \pm 0.07$ & $2.33 \pm 0.08$ & 0.80 & \\ 
HCC-082 & 10:36:33.5 & -27:26:59.5 & $21.59 \pm 0.04$ & $0.77 \pm 0.04$ & $25.09 \pm 0.04$ & $3.41 \pm 0.09$ & $26.55 \pm 0.10$ & $4.64 \pm 0.28$ & 0.61 & \\ 
HCC-083 & 10:36:27.4 & -27:31:27.8 & $21.66 \pm 0.04$ & $0.97 \pm 0.06$ & $23.62 \pm 0.06$ & $1.34 \pm 0.04$ & $25.29 \pm 0.05$ & $2.12 \pm 0.05$ & 0.98 & \\ 
HCC-084 & 10:36:28.0 & -27:30:55.1 & $21.68 \pm 0.04$ & $0.94 \pm 0.05$ & $22.88 \pm 0.04$ & $0.75 \pm 0.02$ & $24.71 \pm 0.05$ & $1.28 \pm 0.03$ & 0.88 & \\ 
HCC-085 & 10:38:11.9 & -27:35:46.9 & $21.70 \pm 0.05$ & $1.01 \pm 0.07$ & $24.31 \pm 0.06$ & $1.89 \pm 0.07$ & $26.15 \pm 0.11$ & $3.23 \pm 0.23$ & 0.79 & \\ 
HCC-086 & 10:36:37.9 & -27:11:40.7 & $21.71 \pm 0.02$ & $0.87 \pm 0.03$ & $23.21 \pm 0.03$ & $0.97 \pm 0.01$ & $25.01 \pm 0.02$ & $1.62 \pm 0.01$ & 0.91 & \\ 
HCC-087 & 10:36:39.0 & -27:21:25.5 & $21.75 \pm 0.05$ & $1.07 \pm 0.08$ & $26.28 \pm 0.05$ & $7.75 \pm 0.51$ & $28.09 \pm 0.36$ & $13.65 \pm 4.62$ & 0.95 & \\ 
HCC-088 & 10:36:36.6 & -27:12:41.2 & $21.76 \pm 0.10$ & $0.84 \pm 0.12$ & $24.36 \pm 0.04$ & $1.70 \pm 0.05$ & $26.20 \pm 0.19$ & $2.79 \pm 0.31$ & 1.13 & \\ 
HCC-089 & 10:36:57.3 & -27:35:13.2 & $21.76 \pm 0.07$ & $0.68 \pm 0.08$ & $24.15 \pm 0.09$ & $1.34 \pm 0.07$ & $25.94 \pm 0.11$ & $2.28 \pm 0.12$ & 0.86 & \\ 
HCC-090 & 10:36:36.4 & -27:35:41.1 & $21.88 \pm 0.02$ & $0.86 \pm 0.03$ & $22.91 \pm 0.05$ & $0.76 \pm 0.02$ & $24.72 \pm 0.03$ & $1.28 \pm 0.02$ & 0.75 & \\ 
HCC-091 & 10:36:52.1 & -27:31:46.7 & $21.88 \pm 0.01$ & $0.93 \pm 0.01$ & $22.70 \pm 0.12$ & $0.79 \pm 0.01$ & $24.73 \pm 0.02$ & $1.47 \pm 0.01$ & 0.50 & \\ 
HCC-092 & 10:36:28.0 & -27:29:36.7 & $21.96 \pm 0.03$ & $0.68 \pm 0.04$ & $24.91 \pm 0.04$ & $2.00 \pm 0.06$ & $26.39 \pm 0.16$ & $2.80 \pm 0.28$ & 0.69 & \\ 
HCC-093 & 10:36:56.5 & -27:30:24.8 & $22.00 \pm 0.03$ & $0.93 \pm 0.06$ & $23.17 \pm 0.06$ & $0.86 \pm 0.03$ & $24.93 \pm 0.04$ & $1.41 \pm 0.03$ & 0.65 & \\ 
HCC-094 & 10:36:22.2 & -27:33:08.5 & $22.01 \pm 0.06$ & $0.75 \pm 0.07$ & $24.71 \pm 0.04$ & $1.93 \pm 0.05$ & $26.46 \pm 0.09$ & $3.11 \pm 0.12$ & 0.91 & \\ 
HCC-095 & 10:36:32.8 & -27:15:15.2 & $22.01 \pm 0.12$ & $0.72 \pm 0.15$ & $24.59 \pm 0.09$ & $1.59 \pm 0.07$ & $26.16 \pm 0.11$ & $2.33 \pm 0.13$ & 0.73 & \\ 
HCC-096 & 10:36:35.4 & -27:15:20.7 & $22.13 \pm 0.04$ & $0.91 \pm 0.07$ & $24.14 \pm 0.05$ & $1.37 \pm 0.04$ & $25.77 \pm 0.08$ & $2.11 \pm 0.07$ & 0.72 & \\ 
HCC-097 & 10:36:53.8 & -27:31:36.3 & $22.14 \pm 0.06$ & $0.68 \pm 0.06$ & $24.45 \pm 0.05$ & $1.42 \pm 0.05$ & $26.49 \pm 0.49$ & $2.68 \pm 0.97$ & 1.10 & \\ 
HCC-098 & 10:36:49.9 & -27:19:46.7 & $22.15 \pm 0.09$ & $0.65 \pm 0.11$ & $24.07 \pm 0.10$ & $1.14 \pm 0.06$ & $25.42 \pm 0.05$ & $1.49 \pm 0.04$ & 0.75 & \\ 
HCC-099 & 10:36:24.6 & -27:22:09.3 & $22.16 \pm 0.10$ & $1.08 \pm 0.14$ & $25.05 \pm 0.11$ & $1.82 \pm 0.11$ & $26.54 \pm 0.13$ & $2.53 \pm 0.19$ & 0.70 & \\ 
HCC-100 & 10:36:25.0 & -27:21:58.1 & $22.24 \pm 0.18$ & $1.03 \pm 0.25$ & $25.07 \pm 0.11$ & $2.00 \pm 0.12$ & $26.28 \pm 0.22$ & $2.26 \pm 0.35$ & 0.53 & \\ 
HCC-101 & 10:37:39.5 & -27:34:05.8 & $22.27 \pm 0.04$ & $1.10 \pm 0.08$ & $23.33 \pm 0.06$ & $0.84 \pm 0.03$ & $25.15 \pm 0.06$ & $1.40 \pm 0.04$ & 1.08 & \\ 
HCC-102 & 10:36:24.5 & -27:14:06.6 & $22.52 \pm 0.19$ & $0.69 \pm 0.24$ & $25.39 \pm 0.09$ & $2.12 \pm 0.11$ & $27.36 \pm 0.23$ & $3.82 \pm 0.47$ & 1.14 & \\ 
HCC-103 & 10:36:38.0 & -27:35:39.8 & $22.55 \pm 0.04$ & $0.98 \pm 0.04$ & $24.60 \pm 0.05$ & $1.99 \pm 0.07$ & $26.39 \pm 0.18$ & $3.35 \pm 0.42$ & 0.83 & \\ 
HCC-104 & 10:38:07.0 & -27:32:03.5 & $22.59 \pm 0.08$ & $0.87 \pm 0.09$ & $24.40 \pm 0.05$ & $1.24 \pm 0.04$ & $25.99 \pm 0.14$ & $1.84 \pm 0.14$ & 0.71 & \\ 
HCC-105 & 10:37:58.5 & -27:31:16.7 & $22.62 \pm 0.12$ & $0.78 \pm 0.14$ & $25.08 \pm 0.05$ & $1.65 \pm 0.06$ & $26.80 \pm 0.32$ & $2.63 \pm 0.57$ & 0.87 & \\ 
HCC-106 & 10:36:39.7 & -27:32:41.2 & $22.64 \pm 0.06$ & $0.96 \pm 0.12$ & $24.52 \pm 0.09$ & $1.17 \pm 0.07$ & $26.38 \pm 0.74$ & $2.01 \pm 1.52$ & 1.02 & \\ 
HCC-107 & 10:37:52.4 & -27:32:19.5 & $22.70 \pm 0.06$ & $0.81 \pm 0.13$ & $24.81 \pm 0.05$ & $1.63 \pm 0.06$ & $26.74 \pm 0.17$ & $2.88 \pm 0.30$ & 1.11 & \\ 
HCC-108 & 10:36:35.5 & -27:13:35.9 & $22.78 \pm 0.14$ & $0.82 \pm 0.22$ & $24.58 \pm 0.08$ & $1.07 \pm 0.05$ & $26.02 \pm 0.08$ & $1.48 \pm 0.06$ & 0.47 & \\ 
HCC-109 & 10:36:24.4 & -27:35:30.7 & $23.01 \pm 0.12$ & $0.73 \pm 0.16$ & $24.62 \pm 0.13$ & $1.01 \pm 0.08$ & $26.31 \pm 0.26$ & $1.60 \pm 0.24$ & 0.73 & \\ 
HCC-110 & 10:36:27.5 & -27:32:12.3 & $23.06 \pm 0.09$ & $0.83 \pm 0.11$ & $25.17 \pm 0.07$ & $1.84 \pm 0.09$ & $26.63 \pm 0.93$ & $2.56 \pm 4.82$ & 0.63 & \\ 
HCC-111 & 10:36:47.3 & -27:13:30.3 & $23.36 \pm 0.16$ & $0.95 \pm 0.22$ & $24.48 \pm 0.17$ & $0.87 \pm 0.61$ & $26.57 \pm 0.52$ & $1.63 \pm 0.29$ & 0.74 & \\

\end{longtable}
\noindent
$^a$ radial velocity taken from \citet{CZ2003}; $^b$ galaxies showing a two component surface brightness profile, not well fitted by a single S\'ersic law \\
$^c$ NGC 3311; $^d$ NGC 3309; $^e$ NGC 3308; $^f$ NGC 3316
}
}

\end{appendix}

\end{document}